\documentclass[12pt,english]{iopart}
\usepackage{cite}
\usepackage{iopams}  
\usepackage[T1]{fontenc}
\usepackage[latin9]{inputenc}
\usepackage{bm,amsthm}
\usepackage{amssymb}
\usepackage{babel}
\usepackage{mathrsfs}
\usepackage{dsfont}
\usepackage{bbold}
\usepackage{fixltx2e}
\usepackage[symbol*]{footmisc}
\usepackage[pdftex]{graphicx} 
\DeclareGraphicsExtensions{.pdf,.jpg,.png}
\newcommand{\Ksh}{\mathcal{K}^{\#}}
\def\num#1{$\left.^{#1}\right.$}
\DefineFNsymbols*{numbers}{{\num0}{\num1}{\num2}{\num3}{\num4}{\num5}{\num6}{\num7}{\num8}{\num9}{\num{10}}{\num{11}}{\num{12}}{\num{13}}{\num{14}}{\num{15}}{\num{16}}{\num{17}}}
\setfnsymbol{numbers}
\renewcommand\mathbb\mathds
\def\RR{\hbox{\bf{\it  I\hskip -2.pt R}}}
\def\CC{\mathbb{C} \kern -.710em \mathbb{C}}
\def\cat#1{{\mathfrak{#1}}}
\newcommand{\bfphi}{\mathbb {\bm\varphi} \kern -.583em \mathbb {\bm\varphi}}
\newcommand{\bfF}{{\bf \mathbb{F}}}
\newcommand{\bfX}{{\bf \mathbb{X}}}
\newcommand{\bfY}{{\bf \mathbb{Y}}}
\newcommand{\bfZ}{{\bf \mathbb{Z}}}
\newcommand{\bfT}{{\bf \mathbb{T}}}
\newcommand{\bfN}{{\bf \mathbb{N}}}
\newcommand{\bfQ}{{\mathcal Q} \kern -.760em {\mathcal Q}} 
\newcommand{\Tau}{\mathbb {\tau} \kern -.583em \mathbb {\tau}}
\newcommand{\One}{\mathbb{1}}
\newcommand{\bfk}{\bf k}
\newcommand{\bfs}{\bf s}
\newcommand{\bfp}{\bf p}
\newcommand{\bfx}{\bf x}
\newcommand{\bfy}{\bf y}
\newcommand{\bfz}{\bf z}

\def\sl#1{ #1 \!\!\! / \  }

\def\catj{\cat{j}}

\def\Xstarnabla{{\;}_{X}\!\!\!\!\!{\;}^{\star}\nabla}

\newcommand{\curlA}{\mathscr{A}\kern -1.015em \mathscr{A}}
\newcommand{\curlD}{\mathscr{D}\kern -.860em \mathscr{D}}
\newcommand{\bfsqcup}{\sqcup \kern -.860em \sqcup}
\newcommand{\bfsqcap}{\sqcap \kern -.860em \sqcap}
\begin{document}
\vspace*{-3cm}	
\title[Functional Integration on Manifold]{Functional Integration on Paracompact Manifolds}
\eqnobysec
\author{Pierre Grang\'e}
\address{Laboratoire Univers et Particules,\\ Universit\'e Montpellier II, CNRS/IN2P3, 
Place E. Bataillon\\ F-34095 Montpellier Cedex 05, France\\
\ead{\mailto{pierre.grange@umontpellier.fr}}}
\author{Ernst Werner}
\address{Institut f$\ddot u$r Theoretische Physik, Universit$\ddot a$t Regensburg,\\ 
Universit$\ddot a$tstrasse 31, \ D-93053 Regensburg,  Germany\\
\ead{\mailto{Ernst.Werner@physik.uni-regensburg.de }}}
\par
\par
\hspace*{1.6cm}\dotfill
\vspace*{-0.4cm}
\begin{abstract}
In 1948 Feynman introduced functional integration. Long ago the problematic
aspect of measures in the space of fields was overcome with the introduction of
volume elements in Probability Space, leading to stochastic formulations. More
recently Cartier and DeWitt-Morette focused on the definition of a proper
integration measure and established a rigorous mathematical formulation of
functional integration. Their central observation relates to the distributional nature
of fields, for it leads to the identification of distribution functionals with Schwartz
space test functions as density measures. This is just the mathematical content of the
Taylor-Lagrange Scheme developed by the authors in a recent past. In this scheme
fields are living in metric Schwartz-Sobolev spaces, subject to open coverings with
subordinate partition-of-unity test functions. In effect these PU, through the
convolution operation, lead to smooth field functions on an extended Schwartz
space. In this way the basic assumption of differential geometry -that fields live on
differentiable manifolds- is validated. Next it is shown that convolution in the
theory of distributions leads to a sound definition of Laplace-Stieltjes transforms,
stemming from the existence of an isometry invariant Hausdorff measure in the
space of fields. Turning to gauge theories the construction of smooth vector fields
on a curved manifold is established, as required for differentiable vector fibre
bundles. The proper choice of a connection further separates in the functional
Hausdorff integration measure a finite Gaussian integration over the gauge
parameter which factors out and plays no physical role. For non-Abelian Yang-
Mills gauge theories with the Vilkowisky-DeWitt connection and the Landau-DeWitt
covariant back-ground gauge result in very simple calculation rules.
\end{abstract}
\vspace*{-0.4cm}
\hspace*{2.3cm}\dotfill
\maketitle
\section{Introduction}
Despite its heuristic nature the formalism of functional integration introduced by Feynman \cite{RPFey} has been implemented in most surveys on Quantum Field Theory (QFT) (e.g.\cite{Ramon,RYD,Hatf}), for it leads to  vacuum-to-vacuum  physical amplitudes of correlation  functions through simple manipulations on the generating functional. The essential mathematical difficulty with this early presentation resides in the naive and elusive definition of an infinite volume element in the space of fields. However rigorous methods  exist to circumvent this problematic. One of the first steps in this direction was through the introduction of volume elements in Probability Space, leading to  stochastic formulations of functional integration \cite{Thooft}. More recently a rigorous  mathematical foundation of functional integration was provided by Cartier and DeWitt-Morette \cite{CardeW}. Given a finite-dimensional vector space $V$, with dual  space $V'$, its central assumption is to consider fields as {\it smooth ${\mathcal C}^\infty$ function} on an extended Schwartz space $\Sigma(V)$, consisting of the Fourier-Stieltjes transforms on $V'$, with a bounded Radon measure $d\mu(z) \in V'$, of an initial field state in the Schwartz space $S(V)$. Then the functional integration takes place on Spaces of Paths where oscillatory integrals appear with integrands of the form $e^{\imath \pi Q(x)}$, where $Q(x)$ is a real quadratic form.  These integrals are made {\it absolutely integrable} ( e.g. on the plane $\RR^2$)  and independent of the necessary regularising functions. The distributional nature of fields in the sense of L. Schwartz \cite{LSCW} is essential for this property.
In the context of Finite Field Theories we have proposed a construction of Quantum Fields as Operator Valued Distributions (OPVD) \cite{JPHYSA1} which leads to  fields as regular ${\mathcal C}^\infty$ quantities. We have successfully applied this formalism to various problems \cite{PGEW05,PGJFMEW,TLRSJM}. In the present paper it is our purpose to incorporate our OPVD-construction into functional integration along the line of Cartier-DeWitt-Morette.
In section $2$ we give a summary of the OPVD formulation in flat space \cite{JPHYSA1}. Starting from the definition of Physical Fields in terms of
bare fields convoluted with fast decreasing PU test functions on Schwartz Spaces $\mathcal{S}(\RR)$ we progress to the introduction of the Path  
Integral Gaussian measure from a generating functional $\chi(\RR^n)$ of a general positive Borel measure on  
$\mathcal{S}(\RR^n)$. Then a first example of functional integration is given for the free scalar field built with PU test functions. In section $3$ we focus on the general setting of functional integration in the context of gauge theories. Due to their geometric characters specific considerations have to be developed for curved manifolds. They encompass, on the one hand, the construction of a smooth  gauge field through a generalisation of the notion of translation in flat space in the distributional context  as flows along geodesics; on the other hand the  separation of the tangent space into horizontal and vertical sub-spaces and the reduced Gaussian integrator is another essential point of the presentation.
In section $4$ special attention is payed to the analysis of Abelian and non-Abelian cases encountered in  contemporary surveys leading to and possibly beyond the Standard Model of fundamental interactions. Finally in the last section some concluding remarks are formulated, pointing out their possible relevance in gravitation viewed as a gauge theory according to \cite{ParTom}.
\section{\bf From fields as OPVD in flat space to the Path Integral with  Gaussian measure on Schwartz Spaces }
\subsection{\bf Summary of the OPVD formulation in Euclidean flat space  $\RR^n$} 
We only quote here the necessary OPVD  background and refer to \cite{JPHYSA1} for a more complete account.\\
Let $\Omega$ be an open subset of $\RR^n$. The genuine topological vector space ${\mathcal D}(\Omega)$ 
of \cite{LSCW} is the space of ${\mathcal C}^\infty$  test functions $\rho$ such that supp 
$\rho \subset \Omega$. To deal with Fourier transforms it is necessary to enlarge the topological 
space ${\mathcal D}(\doteq{\mathcal D}(\RR^n)$ and similar notations in other cases) of 
test functions with compact support to the space  ${\mathcal S}$ of ${\mathcal C}^\infty$ 
test functions $\rho$  of rapid decrease. The dual space of ${\mathcal S}$, 
${\mathcal S}^\prime$, is a subspace of ${\mathcal D}^\prime$:  ${\mathcal S}^\prime \subset {\mathcal D}^\prime$. 
${\mathcal S}^\prime$  is the space of the so-called  {\it tempered} distributions.\\ 
Tempered distributions are of central interest in any QFT formulation. 
 The Fourier transform  ${\mathcal F}[T]$ of a tempered distribution $T$
 is defined by its action on the test function $\rho \in {\mathcal S}(\RR^n)$
 $$<{\mathcal F}[T],\rho>\stackrel{\rm def}{=}<T,{\mathcal F}[\rho]>.$$
 This transform of the test function (resp. tempered distribution) corresponds to an isomorphism of 
 ${\mathcal S}(\RR^n)$  (resp.  ${\mathcal S}^\prime(\RR^n)$)  to itself.
 We denote  ${\mathcal F}[T]$ as $\tilde{T}$ and  ${\mathcal F}[\rho]$ as $\tilde{\rho}$.\\
  To a generalized-field-function  (e.g. distribution) $\phi$ is then associated a continuous linear functional ${\mathcal S}(\RR^n) \ni \rho  \mapsto \Phi[\rho]$
\begin{equation} \label{eq21}
\Phi[\rho]\equiv<\phi,\rho>=\int_{\RR^{^n}} d^{n}y\phi({y})\rho({y}),
\end{equation}
where a smooth flat manifold covered by a single coordinate system (chart) is assumed\footnote{ For a locally integrable $\phi({y})$ the continuous functional $\Phi[\rho]$ defines a finite number, e.g. $\mu$, in the sense of the Theory of Measures. $\mu$ is a finite measure on ${\RR}$ with density $\rho$, for the integral is bounded since $\rho \in {\mathcal D}$ ({\it cf} Appendix A )}.
For $ x \in \RR^n$ the translation operation $\tau_{x}$ defines the translated distribution
 $\tau_{x}\Phi[\rho]$ according to 
\begin{eqnarray}\label{eq22}
\!\!\!\! \tau_{{x}}\Phi[\rho] & = & <\tau_{{x}}\phi,\rho>\stackrel{\rm def}{=}<\phi,\tau_{-{x}}\rho>=\int_{\RR^{^n}} d^{n}y\phi(y)\rho(y-x).
\end{eqnarray}
This is the value of the convolution $\Phi*\rho$ at the point $x$. This translated distribution
enjoys regularity properties \cite{GOLSE} extensively used in signal analysis. However if the mapping $\RR^n \ni x\mapsto \phi(x)$ corresponds to a regular integrable field-function a regularising sequence of test functions $\{\rho_{j \in {\mathcal I}}\}$  exists which converges, when   $j \rightarrow \infty$, to the reflection symmetric  $\delta^{n}(x)$ and ${\displaystyle \lim_{j\rightarrow \infty}}\Phi*\rho_j =\phi$. To implement this possibility the generic test function $\rho$ must also be symmetric under reflection. Then in the dual space the map  $((\RR^n)^\star) \ni {\bf p} \mapsto \tilde{\rho}({\bf p})$ depends only on the norm $\|{\bf p}\|^2 \equiv p^2$. Hereafter we adopt the notation $\tilde{\rho}({\bf p})\equiv f(p^2)$.  \\
  The isomorphisms mentioned in relation to Fourier transforms permits  to rewrite the  
 convolution product in Eq.(\ref{eq22}) as an integral in Fourier-space variables (Parseval-Plancherel relation)
 \begin{eqnarray} \label{eq23}
 (\Phi*\rho)(x)&=&<\tilde{\Phi},e^{(-i\ll .,x\gg)} f>=\int_{\RR^{^n}}\frac{d^np}{(2 \pi)^n} e^{-i\ll {p},{x}\gg}\tilde{\phi}(p)f(p^2).
 \end{eqnarray}
 \par
 It is shown in \cite{JPHYSA1} that for all tempered distributions $\phi \in {\mathcal S}^\prime(\RR^n)$  and all test functions $\rho \in {\mathcal S}(\RR^n)$:\\
\hspace*{0.5cm} {\bf a)} the convolution product $\Phi*\rho$ is in the class ${\mathcal C}^\infty$ on $\RR^n$;\\
\hspace*{0.5cm} {\bf b)} the application $\bm\varphi : \RR^n \ni x \mapsto \bm\varphi(x)=(\Phi*\rho)(x)$ defines a function in the class ${\mathcal S}(\RR^n:\CC).$\\
 In a  $4$-dimensional Minkowski flat space ${\mathcal M}$  with coordinates $x=(x^0,x^1,x^2,x^3)=
(x^0,{\bf x}) \, $ (with a pseudo-metric $g_{\mu\nu}\mbox{diag}\{1,-1,-1,-1\}$, with $\{\mu,\nu\}\in \{0,1,2,3\}\,\,$;$\,\,\square=\partial^\mu g_{\mu\nu} \partial^\nu$ and $\ll{x}{,y}\gg= x^\mu g_{\mu\nu} y^\nu$ the associated bilinear form) to the classical  field-function $\phi(x^0,{\bf x})$ is associated a distribution $\Phi[\rho]$ built on test functions $\rho: (\RR \times \RR^3) \ni (x^0,{\bf x}) \mapsto \rho(x^0,{\bf x})  \in {\mathcal S}(\bar{\RR}^4)$ (the completed topological tensor product ${\mathcal S}(\RR)\hat{\otimes}{\mathcal S}(\RR^3)$ \cite{OSTWI}). The translation-convolution product still takes an integral form in Fourier-space variables and writes\\
 \begin{eqnarray*}
 \!\!\!\!\!\!\!\!(\Phi*\rho)(x^0,{\bf x})&=&<\tilde{\Phi},e^{(-i\ll .,x\gg)} f>=\int\frac{dp_0d^3{\bf p}}{(2 \pi)^4} e^{-i\ll {p},{x}\gg}\tilde{\phi}(p^0,
 {\bf p})f(p_0^2,{\bf p}^2).
 \end{eqnarray*}\\
On general grounds metric spaces are paracompact entities with locally finite open coverings and their subordinate partition of unity. Then on the Euclidean dual space  $(\RR^4) ^\star$ or on its Minkowskian counterpart $(\RR \times \RR^3)^\star$ the respective test functions $f(p^2)$ and $f(p_0^2,{\bf p}^2)$ can be taken as partitions of unity. Some important consequences
follow:\\
\hspace*{0.5cm} {\bf c)} the map $\bm\varphi : \RR^n \ni x \mapsto \bm\varphi(x)=(\Phi*\rho)(x)$ is independent of the PU-test function;\\
\hspace*{0.5cm} {\bf d)} any additional consecutive convolution  $(\cdots((\bm\varphi*\rho)*\rho)\cdots*\rho)$ does not change the initial $\bm\varphi(x)$, for any power of a PU -that is of the test function f- is another but equivalent PU;\\
\hspace*{0.5cm} {\bf e)} the convolution product is Lorentz invariant  and fulfils  Poincar\'e's commutator algebra.\\ 
Moreover if the distribution $\phi \in {\mathcal S}^\prime(\RR^4)$ is that of a classical scalar field function of mass $m$, then its convoluted counterpart $\bm\varphi : \RR^4 \ni x \mapsto \bm\varphi(x)=(\Phi*\rho)(x)$ satisfies the appropriate Klein-Gordon equation and can be taken as the physical field function. After proper identifications of mass-shell distributions and Fock-space operators, the canonical OPVD QFT expression with PU test functions is given in terms of well-defined integrals in the dual Fourier space. In coupled Abelian or non-Abelian field theories the convoluted  gauge vector and Fermi fields functions obey also the initial field equations provided the translation operation is performed with due account of the relevant gauge connection \cite{PGEW05,PGJFMEW,TLRSJM}. From there on a QFT based on the convoluted canonical OPVD fields is obtained, which is free of divergences at any step of the perturbative expansion in the  coupling constant of interacting physical fields functions..
\subsection{\bf Functional integration with  convoluted OPVD fields}
Path integral methods have long been used as valuable tools in the quantisation of field theories. They were early introduced by R.P. Feynmann \cite{RPFey} in a rather heuristic way and developed later on more formal bases by many authors. We shall adopt here the formulation of Cartier-DeWitt-Morette (CDM) \cite{CardeW,JlaCh}, where extensive references on the subject can be found.
\subsubsection{Some results from the theory of measure: Bochner-Minlos theorem in infinite dimensions}
We saw that the convoluted field-functions $\bm\varphi$ belong to the Schwartz space
${\mathcal S}$ of ${\mathcal C}^\infty$-functions with fast decay at infinity. This space ${\mathcal S}$ has a standard  (nuclear \cite{Groten,Trev}) topology and its elements have regularity properties both for small and large distances. There is a well established \cite{GOLSE} natural rigging, or {\it Gel'fand triple}, between ${\mathcal S},\mbox{L}^2$ and the dual space of tempered distributions ${\mathcal S}^\prime, \quad {\mathcal S}\subset \mbox{L}^2 \subset {\mathcal S}^\prime.$
In other words ${\mathcal S}^\prime$, the topological dual of ${\mathcal S}$,
is the space of all continuous linear functionals of $\bm\varphi$ on ${\mathcal S}$. We shall denote $ ^a{\mathcal S}^\prime$ the space of non-continuous linear functionals on ${\mathcal S}$.\\ 
{\bf Theorem 2.2.1} (Bochner-Minlos \cite{Minl,Hida,HidHit,HidKuo}) Let $\chi$ be a function defined on $ {\mathcal S}$ with the following properties,\\
 $(i)$  Normalisation: $\chi(0)=1$;\\
 $(ii)$  Continuity: $\chi$ is continuous on any finite dimensional subspace of ${\mathcal S}$;\\
 $(iii)$ Positivity: $\sum_{k,l=1}^{m}c_k{\bar c}_l\chi(\varphi_k-\varphi_l) \geq 0$, for all $m \in \bfN, c_1 \cdots c_m \in \CC$ and a sequence of test-functions $\varphi_1 \cdots \varphi_m \in {\mathcal S}$.\\
Let ${\mathcal B}(^a{\mathcal S}^\prime)$ be the completion of the Borel $\sigma$-algebra on $ ^a{\mathcal S}^\prime$ ({\it cf} Appendix {\it A}.$1$). Then a unique (probability) measure $\mu$ exists such that for all $\varphi \in  {\mathcal S}$\\
$$\chi(\varphi)=\int_{^a{\mathcal S}^\prime}e^{i \omega(\varphi)}d\mu(\omega),$$
where $\omega(\varphi)$ denotes a linear (non necessarily continuous) functional on $ ^a{\mathcal S}^\prime$. Demanding in $(ii)$ continuity of $\chi$ with respect to a weaker topology on ${\mathcal S}$ yields a measure  supported on the topological dual ${\mathcal S}^\prime$: then $ \omega(\varphi)$ is just the canonical bilinear pairing between ${\mathcal S}$ and ${\mathcal S}^\prime$:
$\omega(\varphi) = <\omega,\varphi>$. $\chi$ is a generating functional of this measure. \\
{\bf Remark 2.2.1} A direct corollary of Theorem 2.2.1  relates the generating functional  $\chi :\varphi \in {\mathcal S} \rightarrow \RR$ to  $\varphi \mapsto \chi(\varphi)=\exp(-\frac{1}{2}\|\varphi\|^2_{L^2})$, where $\|\varphi\|_{L^2}$ is just the usual norm of $\varphi$ in $L^2$. Then $\chi(0)=1$ and $\chi$ is continuous. To verify positivity, let
$V=$ span$\{\varphi_1,\cdots,\varphi_m\}$ with norm $\|.\|_{L^2}$, and $\varphi_1,\cdots,\varphi_m \in {\mathcal S}$ and let $\mu_{V^\prime}$ be the standard Gaussian measure on $V^\prime$ (\cite{CardeW}1997). Then
\begin{equation}\label{chiV}
\int_{V^\prime} e^{i <\omega,\varphi>}d\mu_{V^\prime}(\omega)= e^{-\frac{1}{2}\|\varphi\|^2_{L^2}},
\end{equation}
for every $\varphi \in V$. Therefore
\begin{eqnarray*}
\sum_{j,k=1}^{m}c_j\chi(\varphi_j-\varphi_k){\bar c}_k&=&\sum_{j,k=1}^{m}\int_{V^\prime} c_j e^{i <\omega,\varphi_j-\varphi_k>}{\bar c}_k d\mu_{V^\prime}(\omega)\\
&=&\int_{V^\prime} \mbox{\textbar} \sum_{j=1}^{m} c_j e^{i <\omega,\varphi_j>}\mbox{\textbar}^2 d\mu_{V^\prime}(\omega)
\geq 0 \nonumber
\end{eqnarray*}
{\bf Theorem 2.2.2} (Radon-Nikodyn \cite{Rani}) Let $(V,{\mathcal B},\mu)$ be a finite measure space. If $\nu$ is a finite measure on $(V,{\mathcal B})$, absolutely continuous with respect to $\mu$, then there exists an integrable function $h$ on  $L^1(V,\mu)$ such that $d\nu_E =  h d\mu_E$
for all $E \subset V$. Any two such $h$ are equal almost everywhere with respect
to $\mu$.\\
The function $h \in L^1(V,\mu)$ is dubbed {\it Radon-Nikodyn density} of $\nu$ with respect to $\mu$ and often written as $h=\frac{d\nu}{d\mu}(E)$. Hence $\chi(\varphi)$ is just the Radon-Nikodyn density with respect to $\mu_E(\varphi)$.\\
{\bf Remark 2.2.3} More generally the centred Gaussian measure with mean $\varphi_0 \in E$ and variance $\sigma_{\varphi_0}^2 > 0$ is given by
$$\nu_{\varphi_0,\sigma_{\varphi_0}^2}={\mathcal N}(E,\sigma_{\varphi_0})\int_E \mbox{exp}\left(-\frac{\|\varphi-\varphi_0\|^2}{2\sigma_{\varphi_0}^2}\right) d\mu_E(\varphi),$$
where ${\mathcal N}(E,\sigma_{\varphi_0})$ is a normalisation factor such that, when $\sigma_{\varphi_0} \to 0$, the weak limit of $\frac{d\nu_{\varphi_0,\sigma_{\varphi_0}^2}}{d\mu(\varphi)}(E)$ stands for the degenerate Gaussian (Dirac) measure\footnote{Recall that, by construction, $\varphi \in E \subset {\mathcal S}.$} $\delta_{\varphi_0}$
\begin{equation}\label{fdelta}
\delta_{\varphi_0}\left[\varphi\right]=\int_E\delta_{\varphi_0}\varphi d\mu_E(\varphi)=\varphi_0.
\end{equation}
 Explicitly if an overall atlas $\{(E,\zeta)\}$  with chart function $\zeta$ and coordinates $x,y \in {\zeta[E]}$ is given, we shall have 
$$\delta_{\varphi_0}\left[\varphi\right](\zeta^{-1}(y))=\int_{\zeta[E]}\delta(\varphi(\zeta^{-1}(x))-\varphi_0(\zeta^{-1}(y)))\varphi(\zeta^{-1}(x))\kappa(\zeta^{-1}(x))Dx=\varphi_0(\zeta^{-1}(y)),$$
where $\kappa(\zeta^{-1}(x))Dx=\tilde{\omega}$ is the properly defined volume element on $E$.\\ 
{\bf Remark 2.2.4} (\cite{JAFFE1,DIMO1}) According to our OPVD formalism on a manifold ${\mathcal M}$ with Euclidean metric the test functions are $\rho: \RR^4 \ni x \mapsto \rho(x) \in {\mathcal S}$, which is the {\it projective limit} of an increasing sequence of Sobolev Hilbert spaces \cite{SOBO} $\{{\mathcal H}_s; s \in \bfN\}$ such that the inclusion mapping ${\mathcal H}_{s} \mapsto {\mathcal H}_{s+1}$ is Hilbert Schmidt 
$$ {\mathcal S}=\mbox{{proj-limit}}_{_{_{\hskip -34.pt {s\to\infty}}}}\quad :=   \bigcap_{_{\hskip -2.pt {s\geq0}}}{\mathcal H}_s.$$  
In addition the following equality holds for the dual space 
$$ {\mathcal S^\prime}= \mbox{ind-limit}_{_{_{\hskip -34.pt {s\to\infty}}}}\quad := \bigcup_{_{\hskip -2.pt {s\geq0}}}{\mathcal H}_{-s}.$$
For the free scalar field of positive squared-mass $m^2$ ${\mathcal H}_s$ is a real Hilbert space which can be defined as a completion of $\mathbb{C}^\infty({\mathcal M})$ in the dressed norm
$$\|\rho\|^2_{L^2({\mathcal M}),s} = <\rho,(-\square+m^2)^{-s}\rho>.$$ 
Actually the existence of a first order perturbation to the free scalar field  Hamiltonian $H_0$ may be obtained simply with a test function \ $f({\bf p}^2)\equiv{\mathcal F}[\rho(x)] \in {\mathcal H}_{-1}({\mathcal M})$ ({\it cf} sec. 1.3 of \cite{JAFFE1}). However  perturbations of arbitrary higher orders necessitate test functions $f({\bf p}^2) \in  {\mathcal H}_{-s}({\mathcal M})$ with appropriate values of $s>1$, hence the TLRS formalism developed in \cite{JPHYSA1}.\\ 
{\bf Remark 2.2.5} From what precedes consider now the following situation. Let ${\mathcal H}$ be a Hilbert space identified to its dual. Let ${\mathcal V^\prime}$ be a locally convex real vectorial topological space such that:\\
- ${\mathcal V^\prime}$ is the dual of a space ${\mathcal V}$,\\
- ${\mathcal V}$ is a dense sub-space of ${\mathcal H}$ and the injection ${\mathcal V} \to {\mathcal H}$ is continuous.\\
These spaces form the hilbertian triple ${\mathcal V} \subset {\mathcal H}\subset {\mathcal V^\prime}$, with $<.,.>_{\mathcal H}$ and $\|.\|_{\mathcal H}$  the scalar product and the norm in the Hilbert space ${\mathcal H}$.
The generating Gaussian functional is given by Eq.(\ref{chiV}) and is commonly used to defined the white noise in relation with the standard brownian motion  with ${\mathcal V}= {\mathcal S}(\RR^d)$,${\mathcal H}=L^2(\RR^d)$ and ${\mathcal V}^\prime={\mathcal S}^\prime(\RR^d)$ \cite{HidKuo}. Assuming ${\mathcal H}$ separable and a given overall atlas $\{({\mathcal V},\zeta)\}$ with chart function $\zeta$, there exist an orthonormal numerable basis $(e_i)_{i\in\bfN} \subset {\mathcal V}$ of ${\mathcal H}$ and a dual basis $( ^{\star}e^{j})_{j\in\bfN} \subset {\mathcal V}^\prime$ such that $<e_i\,,^{\star}\!e^{j}>=\delta_i^{\,j}$. Then \\
$$\forall \varphi \in  {\mathcal V}\quad \mbox{and}\quad \omega \in  {\mathcal V}^\prime,\quad
 <\omega,\varphi>_{\mathcal H}={\displaystyle \sum_{j=0}^\infty} <\omega, ^{\star}e^{j}>_{\mathcal H}<e_j,\varphi>_{\mathcal H} \equiv {\displaystyle \sum_{j=0}^\infty}\alpha_j\beta^j,$$
the series converging in $L^2({\mathcal V}^\prime,\mu_{{\mathcal V}^\prime})$. $\chi(\varphi)$ may be written as
$$\chi(\varphi)\equiv\chi(\alpha_1,\cdots\cdots)=\int_{{\mathcal V}^\prime}\mbox{exp}(i{\displaystyle \sum_{j=0}^\infty}\alpha_j\beta^j)d\mu_{{\mathcal V}^\prime}(\beta_1,\cdots\cdots).$$ Of course changing the basis will leave the generating Gaussian functional unchanged. Because of the factorisation of the exponential term in the last integral into an infinite product ${\displaystyle \prod_{j=0}^\infty} e^{(i\alpha_j\beta^j)}$ it would be tempting to relate the Gaussian measure $d\mu_{\mathcal V}(\varphi)$ to the Lebesgue measure ${\displaystyle \prod_j} d\alpha_j$, but the expression is meaningless when $j \to \infty$ \cite{CardeW} (1997, sect.$3.4-3.5$). Thus one has to stick to the generic notation $d\mu_{\mathcal V}(\varphi)$ for the measure.  \\
 To describe a curved base space ${\mathcal M}$ several coordinate chart functions are usually necessary. The OPVD formulation is then far more elaborate as it demands  extensive use of differential geometry and algebraic notions \cite{BarFred}.
 \subsubsection{CDM's scheme for functional integral over spaces of fields} 
The existence of the Gaussian measure on ${\mathcal S}$ being established
the final formulation of the functional integral over the convoluted fields
$\varphi \in {\mathcal S}$ needs further specifications. They are given in full in \cite{BCdeW,Lachap2} and we only indicate in Appendix B the necessary  steps for  our purposes. \\
The Gaussian integrator defined there is ${\mathcal D}_{\Theta,Z}(\varphi)$  on ${\mathcal S}$
with ${\mathcal S}^\prime \ni \Phi \to \Phi*\rho=\varphi$ which can be expressed on ${\mathcal S}^\prime$ in terms of $\phi$ only without alteration of the Gaussian characterisation, for the convolution operation corresponds to a linear change of variable. Moreover the infinite-dimensional determinant of Eq.(\ref{Det}) satisfies the criteria of \cite{CardeW}(1997, sect.$4.1-4.2$) valid for the nuclear translation operator ({\it cf} Appendix C) on the test function $\rho \in {\mathcal S}$.\\
\subsection{Schwinger generating functional for the $n$-point functions of a real massive scalar field as OPVD}
The expression corresponding to Eq.(\ref{TZWeq}), with the usual QFT notation and Fourier conventions  of \cite{Ramon} for the field  $\bm\varphi$ and  for the source distribution $J$, is
\begin{equation} \label{Zphieq}
Z(J)=\int_{{\mathcal S}^\prime}{\mathcal D}_{\Theta,Z}(\bm\varphi)\Theta(\bm\varphi,J),\quad \mbox{where}\quad\Theta(\bm\varphi,J)=\mbox{e}^{\imath(S_0(\bm\varphi)+<J*\rho,\bm\varphi>)},
\end{equation}
with the free action\footnote{It is understood here that with the notation of \cite{Ramon} the squared-mass $m^2$ stands for $m^2-\imath \epsilon$}
\begin{eqnarray}
\fl &S_0(\bm\varphi)=&-\frac{1}{2}\int d^4\bfx\;\bm\varphi(\bfx)\left[\partial_\mu\partial^\mu+\it{m}^2\right]\bm\varphi(\bfx)\nonumber\\
\fl &\;\equiv&\frac{1}{2}\int d^4\bfp\;\it{f}^2(\it{p}_0^2,\overrightarrow{p}^2)\; \tilde{\phi}(\bfp)\;(\bfp^2-\it{m}^2)\;\tilde{\phi}(-\bfp),\\
\fl \mbox{and} \nonumber\\
\fl <J*\rho,\bm\varphi>&\;=&\int d^4\bfx\;\left[{J}*\rho\right](\bfx)\bm\varphi(\bfx)\equiv\int d^4\bfp\;\it{f}^2(\it{p}_0^2,\overrightarrow{p}^2)\tilde{J}(\bfp)\tilde{\phi}(-\bfp).
\end{eqnarray}
Then with the change 
$$\tilde{\phi}^\prime(\bfp)=\left[\tilde{\phi}(\bfp)+\frac{\tilde{J}(\bfp)}{(\bfp^2-\it{m}^2)}\right]\it{f}(\it{p}_0^2,\overrightarrow{p}^2)$$
$Z(J)$ reduces to Schwinger generating functional {\it with PU test-functions} for the $n$-point functions of the free massive scalar field 
\begin{eqnarray}\label{ZJms}
Z(J)&=&{\mathcal N}\int {\mathcal D}_{\Theta,0}(\bm\varphi^\prime)\mbox{exp}\left[\frac{\imath}{2}\int d^4\bfx\;\bm\varphi^\prime(\bfx)\left[\partial_\mu\partial^\mu+\it{m}^2\right]\bm\varphi^\prime(\bfx)\right] \nonumber\\
& &\times\mbox{exp}\left[-\frac{\imath}{2}\int d^4\bfp\;\frac{\mid\tilde{J}(\bfp)\mid^2\it{f}^2(\it{p}_0^2,\overrightarrow{p}^2)}{\bfp^2-\it{m}^2}\right]\nonumber,\\ &=&\mbox{exp}\left[-\frac{\imath}{2}\int d^4\bfp\;\frac{\mid\tilde{J}(\bfp)\mid^2\it{f}^2(\it{p}_0^2,\overrightarrow{p}^2)}{\bfp^2-\it{m}^2}\right]
\end{eqnarray}
When the free action is extended to include an interaction term, $S_{int}$,  perturbative contributions in its coupling strength are obtained in the usual way   through successive higher order functional derivatives with respect to the source $J$ ({\it cf e.g.} \cite{Ramon}), with however the presence of PU test-functions on every internal propagating lines.
\setcounter{equation}{0}
\section{ OPVD gauge-field theories: geometrical considerations and Gaussian integrators}
\subsection{OPVD formulation and gauge transformations for the gauge field}
The naive path integral suggested by the scalar field theory points to integration over all configuration of the gauge field, {\it e.g.} $A_\mu$, as a distribution. However, in line with  Sec.2.1, the gauge field operator functional in flat space is now
\numparts
\begin{eqnarray} \label{eqArho}
\tau_{x}{\bf A}_{\mu}[\rho]&=&<\tau_{x}A_{\mu},\rho> = <A_{\mu},\tau_{-x}\rho> \equiv {\bf A}_{\mu}[\rho](x)\\ 
&=&\int d^{4}y A_{\mu}(y) \rho(y-x).
\end{eqnarray}
\endnumparts
In the sequel, unless otherwise stated, for ease of notation we shall write ${\bf A}_{\mu}(x)$ instead of ${\bf A}_{\mu}[\rho](x)$. 
In the abelian case it easy to see that ${\bf A}_{\mu}(x)$ transforms as the original $A_{\mu}(y)$ under a gauge transformation \cite{MGW}.
In the Yang-Mills case let $v=v^\alpha{\bf T}_\alpha$ be an infinitesimal gauge transformation parameter,$\,\,U=1-v$, with Lie group generators $\{{\bf T}_\alpha\}$. The original gauge field\footnote{$\mathscr{A}=\mathscr{A}_\mu dx^\mu=A^{\alpha}_{\;\;\mu}{\bf T}_\alpha dx^\mu$ within the given coordinate system} $\mathscr{A}_\mu$ transforms as
$$\mathscr{A}_\mu \rightarrow (1+v)(\mathscr{A}_\mu+d)(1-v)=\mathscr{A}_\mu-\mathscr{D}_\mu v,$$
where $d$ is the exterior derivative of usual calculus on manifolds and $\mathscr{D}_\mu v=\partial_\mu v +[\mathscr{A}_\mu,v]$. Assuming for simplicity a single chart over the whole flat manifold the transformed convoluted gauge field ${\;}^{U}\!\!{\curlA_\mu}$ writes
\begin{eqnarray*}
{\;}^{U}\!\!{\curlA_\mu}&=&\curlA_\mu+\int d^4y\; v(y)\; \mathscr{D}_\mu\;\rho(y-x)\\
&=&\curlA_\mu-\curlD_\mu{\bf v}+\mathscr{O}(v^2),
\end{eqnarray*}
where $\curlD_\mu {\bf v}= \partial_\mu {\bf v} +[\curlA_\mu,{\bf v}]$, with ${\bf v}$ the convoluted gauge transformation parameter. This is in the most direct correspondence to the abelian case. The generalisation of the convolution operation to curved manifolds is postponed to the end of subsection $3.3.2$.\\
Four-dimensional Abelian and Yang-Mills gauge theories are gauge-invariant but highly divergent in their usual treament  because of the presence  of ill-defined product of distributions at the same space-time points. 
The perturbative treatment of these divergences in a gauge invariant way has long been a major issue. As shown in \cite{MGW} the introduction of the OPVD formalism leads to extensions of singular distributions both in the IR and UV regimes, which are consistent with conservation of gauge symmetry directly at the physical dimension $D=4$. An important aspect of the OPVD formalism resides in the possibility of specific choices of test 
functions, ${\;}^{\perp}\rho(x) \in {\mathcal S}$, {\it horizontal} with respect to the the tangent hyperplane. Thereby the gauge field action is restricted to an expression containing only physically relevant gauge components of ${\bf A}_{\mu}$. The procedure can be exemplified in an elementary way adapted from \cite{Hatf}.
\subsection{A simple example}  
Let $\bfX$ and $\bfY$ be two real separable Banach   spaces with their duals $\bfX^\prime$ and $\bfY^\prime$. The  Gaussian integrator ${\mathcal D}(\bfx)$ (resp.${\mathcal D}(\bfy)$) is characterised ({\it cf}
Appendix B and \cite{CardeW}(1997)) by the following integration formula:
$$\int_{\bfX}{\mathcal D}(\bfx).\exp\left[-\pi\bfQ(\bfx)-2\imath\pi<\bfx,\bfx^\prime>\right]=\exp(-\pi W(\bfx^\prime)) $$
for every $\bfx^\prime \in \bfX^\prime$, $\bfQ(\bfx)>0 $ is a quadratic form in $\bfX$ and ${\bf W}(\bfx^\prime)$ its inverse in the dual space $\bfX^\prime$. Provided the Fourier-Stieltjes transform of the measure $\mu$ on $\bfX^\prime$ in Eq.(\ref{FStr}) is replaced by its Laplace-Stieltjes transform these integrators have the property ({\it cf} Appendix A for details)
$$ {\mathcal D}(\bfx+\bf\Lambda)={\mathcal D}(\bfx), \quad \bf\Lambda\; \mbox{a fixed element of}\; \bfX.$$
Consider the maps
$$Q_{\pm}: \bfX\times\bfY \ni (\bfx,\bfy)\; \stackrel{\rm Q_{\pm}}{\mapsto} \quad Q_{\pm}(\bfx,\bfy)=\exp(-\|\bfx\pm\bfy\|^2).$$
Let us call a gauge transformation the symmetry operation $(\bfx,\bfy) \to (\bfx+\bf\Lambda,\bfy+\bf\Lambda)$ for $Q_{_{-}}(\bfx,\bfy)$. A gauge "orbit" is the path of  a "configuration" $(\bfx,\bfy)$ through $\bfX\times\bfY$ under a gauge transformation, that is a line of constant norm in $\bfx-\bfy$. A gauge invariant "action" $S$ is a map
$$S:\bfX\times\bfY \ni (\bfx,\bfy) \stackrel{\rm S}{\mapsto} S(\bfx-\bfy).$$\\
{\bf Theorem 3.2.1}  Let $\{\bfX,{\mathcal B}(\bfX),\mu\}$ and $\{\bfY,{\mathcal B}(\bfY),\nu\}$ be two  $\sigma$-finite measured spaces. Then:\\
 i) a unique measure $m$ exists on $\{\bfX\times\bfY,{\mathcal B}(\bfX)\otimes{\mathcal B}(\bfY)\}$ such that $m(\bfX\times\bfY)=\mu(\bfX)\nu(\bfY)$\\
ii) for all $E \in {\mathcal B}(\bfX)\otimes{\mathcal B}(\bfY)$ one has:
$$m(E)=\int_{\bfX}\nu(E_{\bfx})d\mu_{\bfx}=\int_{\bfY}\mu(E^{\bfy})d\nu_{\bfy}.$$
The proof can be found in \cite{Integ}.\;\;$\blacksquare$\\
As a consequence and from the link indicated in Eq.(\ref{TZWeq}) the Gaussian integrator on $\bfX\times\bfY$ is just ${\mathcal D}(\bfx){\mathcal D}(\bfy)$.
Switching to the "configuration" $(\bfz_{+},\bfz_{-})$ with $\bfz_{\pm}=\frac{\bfx\pm\bfy}{2}$ we have 
\begin{equation}\label{Zmod}
\fl {\mathcal Z}\!=\!\!\int_{\bfX\times\bfY}\!\!\!\!\!\!\!{\mathcal D}(\bfx,\bfy)Q_{+}(x,y)Q_{-}(x,y)e^{-S(\bfx-\bfy)}\!=\!\!\int\!\!\!{\mathcal D}(\bfz_{+})e^{-\|\bfz_+\|^2}\!\!\int\!\!\!{\mathcal D}(\bfz_{-})e^{(-\|\bfz_{-}\|^2-S(\bfz_{-}))}.
\end{equation}

Actually we can change the "variables" $(\bfx,\bfy)$ to $(\bfz_{-},\Lambda)$ instead of $(\bfz_{-},\bfz_{+})$, where ${\bf\Lambda}$ is the gauge transformation that takes a configuration point on the line $\bfz_{+}=\bfx+\bfy={\bf 0}$ to $(\bfx,\bfy)$ on the gauge orbit\footnote{The gauge dependent variable $\bfz_{+}$ is actually traded for the gauge transformation that brings $\bfz_{+}$ from ${\bf 0}$ to its position.}. Then ${\mathcal Z}$ writes  
$${\mathcal Z}=\int{\mathcal D}(\Lambda)\;e^{-\|\Lambda\|^2}\int{\mathcal D}(\bfz_{-})\;e^{(-\|\bfz_{-}\|^2-S(\bfz_{-}))},$$
where the gauge contribution factors out explicitly. These integrals can be written once more as
 $${\mathcal Z}=\int{\mathcal D}(\Lambda)\;e^{-\|\Lambda\|^2}\int{\mathcal D}(\bfx)
{\mathcal D}(\bfy)2\delta_{{\bf 0}}(\bfx+\bfy)\;e^{-\|\bfx + \bfy\|^2}\;e^{-(\|\bfx - \bfy\|^2+S(\bfx - \bfy))},$$
with the distribution $\delta$ defined in Eq.(\ref{fdelta}). Its interpretation  in our disguised gauge-language corresponds to a gauge-fixing term. A different gauge choice is specified by a map 
$$\mbox{{\bf f}}: \bfX\times\bfY \ni (\bfx,\bfy)\; \stackrel{\rm \mbox{{\bf f}}}{\mapsto}\;\mbox{{\bf f}}(\bfx,\bfy)={\bf 0}$$
 defining equivalent classes of geometry called the  "gauge slice". A {\it bona fide} gauge choice is such that any two points on the gauge slice are unconnected by the symmetry. As shown in \cite{Hatf} for a given gauge choice $\mbox{{\bf f}}(\bfx,\bfy)={\bf 0}$ the extraction of the gauge contribution due to equivalent classes of geometry is obtained through the introduction of the identity $ \int{\mathcal D}(\mbox{{\bf f}})\delta_{{\bf 0}}(\mbox{{\bf f}})={\bf 1}$ in Eq.(\ref{Zmod}) and comes out as an elementary form of the general Faddeev-Popov algebraic method \cite{FadPop}. \\
On this simple example it is easy to catch the essence of the general alternative geometric approach based on {\it separation}  of tangent spaces into "vertical" and "horizontal" sub-spaces.\\ 
Let us introduce a tensor field $\mathscr{J}$ such that at each point $\bfz$ of ${\mathcal M}_{XY} =\bfX\times\bfY \; \mathscr{J}_{\bfz}^2=\One_{\bfz}$, that is $\mathscr{J}_{\bfz}$ has eigenvalues $\pm 1$. The extraction of equivalent classes of geometry from the composite  space is accomplished
by dividing ${\mathcal M}_{XY}$ into two disjoint sub-spaces according to the eigenvalues of $\mathscr{J}_{\bfz}$, the positive (resp. negative) one corresponding here to the "vertical" (resp. "horizontal") decomposition,
$${\mathcal M}_{XY} ={\mathcal M}_{XY}^{-}\;\oplus\;{\mathcal M}_{XY}^{+}.$$
where,
$${\mathcal M}_{XY}^{\pm}=\{\bfz \in {\mathcal M}_{XY}\;\mid \;\mathscr{P\!\!\!J}^{\pm}_{\bfz}\stackrel{\rm def}{=}\frac{1}{2}(\One\pm\mathscr{J}_{\bfz});\;\mathscr{P\!\!\!J}^{\pm}_{\bfz}.\bfz=\bfz_{\pm}\}.$$ 
It is clear that the above decomposition, that is the definition of the projectors
$\mathscr{P\!\!\!J}^{\pm}_{\bfz}$, is tied up to a proper\footnote{In the sense that  equivalent classes of geometry have been fully taken into account.} choice of gauge. Then the final determination of the generating function ${\mathcal Z}$ in Eq.(\ref{Zmod}) only depends on the quotient space ${\mathcal M}_{XY}\slash \mathcal{G}$, where $\mathcal{G}$ is the gauge-group of transformations. Simply stated an "horizontal" projection of the initial Gaussian measure on ${\mathcal M}_{XY}$ is bound to achieve the necessary work. \\
The observations made here on a simple example gain full interests for Abelian and Yang-Mills gauge fields when treated as OPVD in (nuclear) Schwartz spaces ${\mathcal S}^\prime$, for the horizontally-projected test functions remains in ${\mathcal S}$ ({\it cf} Sec. (2.1-2.2) of \cite{MGW}).
\subsection{Geometrical setting for OPVD gauge fields}
Our objective is to define the gauge theory directly on the space of non-equivalent
gauge potentials {\it e.g.} the orbit space. The geometrical setting to achieve the description of the physical configuration space of the system is given in many text books \cite{Mayer,DanVial,Nakah}. A quick overview introducing the necessary ingredients is presented in Appendix D. 
\subsubsection{Separation of the tangent space into  vertical and horizontal sub-spaces}
 Let ${\mathcal G}$ be a Lie group. $\forall a,h  \in {\mathcal G}$ the right action $R_a$ is defined by $R_ah=ha$ and on the tangent space$\;\;T_p{\bf P} \;\; R_a$ induces a map $R_{a^\star}: T_{p}{\bf P}\mapsto T_{pa}{\bf P}\;\; \forall  p \in {\bf P}$.\\
{\bf Definition 3.3.1} A connection on the principal bundle $\{ {\bf P},{\mathcal G},\Pi\}$ is a unique separation of the tangent space $T_p{\bf P}$ into the vertical 
and horizontal sub-spaces $T^{\parallel}_p{\bf P}$ and $T^{\perp}_p{\bf P}$ such that\\
(i) $T_p{\bf P}= T^{\perp}_p{\bf P} \oplus T^{\parallel}_p{\bf P},$\\
(ii) A {\it smooth} vector field ${\bf A}$ on ${\bf P}$ is separated into smooth vector fields ${\bf A}^{\parallel} \in T^{\parallel}_p{\bf P}$ and  ${\bf A}^{\perp} \in T^{\perp}_p{\bf P}$ with ${\bf A}={\bf A}^{\parallel} + {\bf A}^{\perp}, $\\
(iii) $T^{\perp}_{pa}{\bf P}=R_{a^\star}T^{\perp}_{p}{\bf P}\quad \forall a \in {\mathcal G}$ and $p\in {\bf P}$\\
{\bf Remark 3.3.1} It is only with smooth vector fields that the action of projection operators, combinations of partial differentials, is well defined in the usual sense.
Hovewer the field objects under consideration are not smooth, for they are vector fields distributions. In the present geometrical context it is only after the extension of the convolution operation of sect.($3.1$) that the necessary smoothness is achieved.\\
{\bf Remark 3.3.2} The last condition ensures that if a  point $p \in {\bf P}$ is parallel transported, so is its constant multiple $pa \;\; a \in {\mathcal G}$.
\subsubsection{Connection, covariant derivative, gauge condition, parallel transport and construction of the smooth vector field}
Since ${\mathcal G}$ is a group manifold it admits a group invariant metric, extendable in infinitely many ways to an invariant metric $\bm\gamma$ on the principal fibre bundle.
Under the requirement of ultralocality this extended metric turns out to be unique for Yang-Mills gauge theories \cite{DeWMol}.\\
The classical action $S({\bf A})$ is invariant under infinitesimal  gauge transformations of the form
$$ \delta{\bf A}^i= T^i_\alpha\delta\xi^\alpha,$$
where ${\bf A}^i$ are the fields (coordinates in the principal fibre bundle), $\{{\bf T}_\alpha\}$ the set 
of vector fields with Lie algebra bracket relations that generate the fibre
 and $\delta\xi^\alpha$ the infinitesimal gauge parameters. Here, for  convenience of notation, we adopt DeWitt's condensed index convention \cite{BCdeW}: all indices play a double r\^ole specifying discrete labels as well as space-time points with the meaning that summations over repeated indices include integrations over space-time. Gauge invariance of the metric $\bm\gamma$ is the statement that the vector fields ${\bf T}_\alpha$ with components $T^i_\alpha$ are the Killing vector fields for $\bm\gamma$ and vertical vector fields for the principal bundle \cite{ParTom}, that is in term of the Lie derivative
$$\mathscr{L}_{{\bf T}_\alpha}\bm\gamma=0.$$
The ${\bf T}_\alpha$ and $\bm\gamma$ together define a $1$-form on the principal bundle
$${\bm\omega}^\alpha=\cat{G}^{\alpha\beta}{\bf T}_\beta.\bm\gamma$$
where $\cat{G}^{\alpha\beta}$ is the Green function (with appropriate boundary conditions) of the operator
$$\cat{F}_{\alpha\beta}=-{\bf T}_\alpha.\bm\gamma.{\bf T}_\beta.$$
By construction it is seen that $\bm\omega^\alpha.{\bf T}_\beta\!\!=\!\!\delta^\alpha_{\;\;\beta}$ and that horizontal gauge vectors perpendicular to the fibre (under the metric $\bm\gamma$) are obtained by application of the horizontal projection operator
$$ \Pi^i_{\;\;j}=\delta^i_{\;\;j}-T^i_{\;\;\alpha}\omega^\alpha_{\;\;j}.$$
The covaraint derivative is usually specified in term of the Riemannian connection $\bm\Gamma_\gamma$ associated with the metric $\bm\gamma$. The method proposed by Vilkovisky-DeWitt \cite{Vilk1,Vilk2,DeWgh,DeWittQFT} to avoid ghosts uses the fact that  $\bm\Gamma_\gamma$ is defined up to an arbitrary tensor field.
\paragraph{\;{\it The Vilkowisky-DeWitt connection}}
The metric on the space (horizontal) of distinct gauge orbits, orthogonal to the space of generators of gauge transformations, is often commonly denoted $\bm\gamma^{\perp}$. It is given by $$\gamma^{\perp}_{ij}=\Pi^k_{\;\;i}\Pi^l_{\;\;j}\gamma_{kl}.$$
Let $\bar{\nabla}$ be the covariant derivative with respect to the connection $\bar{\Gamma}^k_{ij}$. Gauge invariance of the metric results in a natural choice for this connection through the requirement that\footnote{here $\gamma^{\perp}_{jk,l}$ stands for $\partial_l\gamma^{\perp}_{jk}$ and alike}
$$\bar{\nabla}_i\gamma^{\perp}_{jk}=0=\gamma^{\perp}_{jk,i}-\bar{\Gamma}^l_{ij}\gamma^{\perp}_{lk}-\bar{\Gamma}^l_{ik}\gamma^{\perp}_{jl}\;\;,$$
which leads to
$$\bar{\Gamma}^l_{ij}\gamma^{\perp}_{lk}=\frac{1}{2}(\gamma^{\perp}_{jk,i}+\gamma^{\perp}_{ki,j}-\gamma^{\perp}_{ij,k}).$$
Since $\gamma^{\perp}_{ij}T^j_{\;\alpha}=0$ the inversion of $\gamma^{\perp}_{lk}$ over the full field space  should lead to a connection $\bar{\Gamma}$ which, besides the usual metric and arbitrary tensor-type terms, involves another arbitrary term, multiple of $T^l_{\;\alpha}$.\\
The procedure for avoiding ghosts is to make use of the following connection \cite{Vilk2} on the frame bundle $F{\bf P}$:
$$\bar{\Gamma}^i_{\;\;jk}={\Gamma_\gamma}^{i}_{\;\;jk}+\cat{V}^i_{\;\;jk}+T^i_{\;\;\alpha}C^\alpha_{\;\;jk}$$
 where $\bm\Gamma_\gamma$ is the Riemannian-Christoffel $\gamma$-metric connection and $C^\alpha_{\;\;jk}$ is completely arbitrary. Vilkovisky's connection proper is 
$$\cat{V}^i_{\;\;jk}=-T^i_{\;\;\alpha.j}\omega^\alpha_{\;\;k}-T^i_{\;\;\alpha.k}\omega^\alpha_{\;\;j}+\frac{1}{2}\omega^\alpha_{\;\;j}T^i_{\;\;\alpha.l}T^l_{\;\;\beta}\omega^\beta_{\;\;k}+\frac{1}{2}\omega^\alpha_{\;\;k}T^i_{\;\;\alpha.l}T^l_{\;\;\beta}\omega^\beta_{\;\;j},$$
where the dots denote covariant differentiation based on the index that follows. This connection has important remarkable properties detailed in \cite{DeWMol}. The first three of them relevant to our purposes are:\\
(i) Let ${\bf c}$ be a geodesic based on the connection $\bar{\bm\Gamma}_\gamma$. If the tangent vector to ${\bf c}$ is horizontal at one point along ${\bf c}$ then: (a) it is horizontal everywhere along ${\bf c}$, (b) ${\bf c}$ is also a geodesic based on $\bm\Gamma_\gamma$, and (c) ${\bf c}$ is the horizontal lift of a Riemannian geodesic in ${\bf P}/{\mathcal G}$ based on the existing natural projection of $\bm\gamma$ down to ${\bf P}/{\mathcal G}$.\\
(ii) Alternatively, if ${\bf c}$ is tangent to a fibre at one point then it lies in that fibre.\\
(iii) For all $\alpha$ $T^i_{\;\;\alpha;j}=\frac{1}{2}T^i_{\;\;\gamma}f^\gamma_{\;\;\alpha\beta}\omega^\beta_{\;\;j}$, where the semicolon denotes covariant functional differentiation based on Vilkovisky's connection and the $f^\gamma_{\;\;\alpha\beta}$ are the ${\mathcal G}$ group stucture constants.\\
The outcome is that in a covariant Taylor expansion of the classical action gauge invariance imposes the annulation of the term in $C^\alpha_{\;\;jk}$ and only the Christoffel connection ${\Gamma_\gamma}^{i}_{\;\;jk}$ and the term $\cat{V}^i_{\;\;jk}$ contribute to the result. Moreover to all order in the expansion the resulting effective action is independent of the choice of gauge and for Yang-Mills theory
in the Landau-DeWitt covariant background gauge \cite{FradTse} there is no contribution from $\cat{V}^i_{\;\;jk}$ either. Explicit calculations \cite{FradTse,BarVilk,RebA,HugKun,TomDJ} did verify these statements. The calculational rules for Yang-Mills theory turned out to be very simple \cite{DeWMol} as every Feynman graph containing a ghost line is found to vanish identically  due to specific forms of ghost-vector-field vertices.\\
What we have exposed so far corresponds to the {\it standard}  geometrical setting
for gauge vector fields. In this framework the  basic assumption of smoothness for these fields is however shaky from the start, for they are not regular functions but distributions. The convolution procedure of Sec.$3.1$ in principle remedies to this native defect but needs specifications in the context of the Vilkovisky-DeWitt connection pointed above. To do so we have first to introduce the notion of {\it local} convolution.
\paragraph{\;{\it Local convolution \newline}}
{\bf Definition 3.3.2} Let $\RR^+ \ni r$, $\Omega^{\catj}_r=\{x:\mbox{ball}\;(x,r) \subset \Omega^{\catj}\}$ and $\alpha^{\catj} \in \mathscr{C}^\infty(\Omega^{\catj})$ a localisation function such that $\alpha^{\catj}\equiv 1$ on $\Omega^{\catj}_r$ with $\mbox{supp}\;\alpha^{\catj} \subset  \Omega^{\catj}$. Take distributions $\Phi \in \mathscr{D}^\prime(\Omega^{\catj})$ and $\Upsilon \in \mathscr{D}^\prime(\RR^4)$ such that $\mbox{supp}\;\Upsilon \subset {\bf B}$, where ${\bf B}=\{x:\mid x \mid < r\}$. Then the local convolution $\Phi\star_{\;_{\hskip -11.pt \mbox{loc}}}\!\!\Upsilon \in \mathscr{D}^\prime(\Omega^{\catj}_{2r})$ is defined by
$$\Phi\star_{\;_{\hskip -11.pt\mbox{loc}}}\!\!\Upsilon=\widetilde{\alpha^{\catj}\Phi}\star\Upsilon|_{\Omega^{\catj}_{2r}},$$
where $\widetilde{\alpha^{\catj} \Phi} \in \mathscr{D}^\prime(\RR^4)$ is the prolongation by $0$ \; (defined by $<\!\widetilde{\alpha^{\catj} \Phi},\rho\!>_{\RR^4}=<\!\Phi,\alpha^{\catj}\rho\!>_{\Omega^{\catj}}$ for all test function $\rho$).\\
{\bf Lemma 3.3.1} The convoluted local distribution does not depend on the choice of  $\alpha^{\catj}$. It is conserved under restriction, that is $\Phi|_\omega\;\;\star_{\!_{\hskip -13.pt {\mbox{loc},\omega}}}\!\!\Upsilon=\Phi\;\;\;\star_{\;_{\hskip -19.pt{\mbox{loc},\Omega^{\catj}}}}\!\!\Upsilon|_{\omega_{2r}}$ and verifies
$$\partial_j(\Phi\star_{\;_{\hskip -11.pt\mbox{loc}}}\!\Upsilon)=\partial_j\Phi\star_{\;_{\hskip -11.pt \mbox{loc}}}\!\!\Upsilon=\Phi\star_{\;_{\hskip -11.pt\mbox{loc}}}\!\!\partial_j\Upsilon.$$
The proof can be found in \cite{Simj}.\;\;$\blacksquare$\\
Clearly these properties are verified if, say $\Upsilon$, is itself taken as a regular test function like $\rho$ in Sec.$3.1$. It was assumed  that each $(\Omega^{\catj})_{j\in J}$ is equiped with a local chart $(\Omega^{\catj},\zeta)$ where points $p,q,\cdots$ have  coordinates $x,y,\cdots \in \zeta(\Omega^{\catj})$. In this way it forms a smooth atlas of  $\Omega=\bigcup_{\catj \in J}\Omega^{\catj}$. 
The family $(\alpha^{\catj})_{\catj \in J}$ of positive continuous functions on $\Omega$ is a  partition of unity (PU) subordinate to $(\Omega^{\catj})_{\catj\in J}$. For a given point $p$ of coordinates $x \in \zeta(\Omega^{\catj}_{2r})$ and for all  points $q$ of coordinates $y \in \zeta({\bf B})$ let ${\bf c}_x^y$ be a geodesic, parametrised by $c: t \in \RR \to c(t) \in \Omega^{\catj}_{2r}$ with  $c(0)=\zeta^{-1}(x),c(t)=\zeta^{-1}(y)$, and based on the connection $\bar{\bm\Gamma}_\gamma$. Then if the distribution $\Phi$ stands for the Lie-algebra-valued one form $\mathscr{A}^{\catj}\in \cat{g}\otimes T_p^\star\Omega^\catj$ associated to that connection, the corresponding local convolution is expressed in terms of the transportation from $p$ to $q$ of the scalar function $\alpha^\catj \rho$ by the flow $f_t \;t\in \RR$, generated by the Killing vector field ${\bf T}_\beta$ for the metric $\bm\gamma$, that is 
$$(\alpha^\catj \rho)_{t}(\zeta^{-1}(y))= \alpha^\catj \rho(f_{-t}(\zeta^{-1}(y))),$$ 
with $\alpha^\catj \rho=(\alpha^\catj \rho)_{t=0}$  and where the flow $f_t$ obeys  the infinitesimal transport equation $\frac{df_t}{dt}=-{\bf T}_\beta(f_t)$ with initial condition $f_0=\mathbb{I}$. Then for $p=\zeta^{-1}(x) \in \Omega^{\catj}_{2r}$

\begin{eqnarray}\label{convol}
\fl &&\;\;\;\curlA^{\catj}_\mu(\zeta^{-1}(x))=\mathscr{A}^{\catj}_\mu\star_{\;_{\hskip -11.pt \mbox{loc}}}\!\!\rho(\zeta^{-1}(x))=\int_{\zeta({\bf B})}\!\!\!\mathscr{A}^{\catj}_\mu(\zeta^{-1}(y))(\alpha^\catj \rho)_{t}(\zeta^{-1}(y))\kappa(\zeta^{-1}(y))Dy,
\end{eqnarray}
where $\kappa(\zeta^{-1}(y))Dy=D_{Vol}({\bf B})$ is the properly defined volume element on ${\bf B}$.\\
{\bf Remark 3.3.3} If the initial $\mathscr{A}^\catj(p)$ is horizontal the convoluted expression $\curlA^{\catj}(p)$ keeps this property, for the metric Killing vector field conserves the geometry along the flow $f_t$. It is clear then, from the second part of Lemma 3.3.1, that using a projected test function ${\;}^{\perp}\!\rho^{i}_{\;\;j}(p)=\Pi^i_{\;\;j}\rho(p)$ picks up the horizontal projection of $\mathscr{A}^\catj(q)$ and gives in turn  the horizontal part of the convoluted smooth vector field $\curlA^{\catj}(p)$.
It is then of interest to investigate first the role of the gauge fixing condition in the determination of the precise form of the corresponding horizontal  projection operator. \\ 
{\bf Remark 3.3.4} Fixing a gauge by $\chi^\alpha(\curlA^{\catj})=0$ amounts to the choice of a local coordinate system which may become singular: this is precisely where the Gribov ambiguity appears \cite{Grib} ({\it cf} \cite{HeinzlP} for a didactic account). The transportation of the scalar function $\alpha^\catj \rho$ by the flow generated by the Killing vector field for the metric implicitly assumes that all geodesics can be prolonged to infinity, forming a {\it geodesically complete orbit space}. That the gauge choice satisfies this property and solves the Gribov problem is usually elaborate.\\
{\bf Proposition 3.3.1} Let $\chi^\alpha(\curlA^{\catj})=0$ be the local gauge-fixing condition of the vector fields $\curlA^{\catj}\;\in\;T_p{\bf P}$, the tangent space built on the fibre bundle ${\bf P}$. Under precise solvability requirements the  gauge-fixing condition induces a modified set of generators of $\cat{g}$ leading to a specific form of the horizontal projector $\bm\Pi$.\\
 {\bf Proof}:  Under an infinitesimal change of gauge $\delta\xi^\alpha$ we require that $\chi^{\alpha} (\curlA^{\catj}+\delta{\curlA^{\catj}})=\chi^{\alpha}(\curlA^{\catj})$ has the unique solution
solution $\delta\xi^\alpha=0$, that is
\begin{equation}\label{Kij}
Q^\alpha_{\;\;\beta}(\curlA^{\catj})\delta\xi^\beta=0\;\;\mbox{with}\;\;Q^\alpha_{\;\;\beta}(\curlA^{\catj})=\chi^{\alpha}_{\;\;,i}(\curlA^{\catj})K^i_{\;\;\beta},
\end{equation}
where $\chi^\alpha_{\;\;,i}(\curlA^{\catj})$ denotes the functional derivative of $\chi^\alpha(\curlA^{\catj})$ with respect to $\curlA^{\catj}$ and $K^i_\alpha$ can be regarded \cite{TomDJ} as the generator of gauge transformations, denoted ${\bf T}^i_\alpha$ in sect. $3.3.1$. Provided that\footnote{we shall see that det$Q^\alpha_{\;\;\beta}$ is just the Faddeev-Popov factor mentioned in our simple example.} det$Q^\alpha_{\;\;\beta}\neq0$  the constraint on  $\chi^\alpha(\curlA^{\catj})$ implies, to lowest order in the change of gauge, the necessary solution $\delta\xi^\alpha=0$. Define then\footnote{The two sets of generators differing by a non-singular {\it real} linear transformation span the same Lie algebra and generate the same group.} $\tilde{{\bf T}}_\beta={\bf T}_\alpha Q^\alpha_{\;\;\beta},\;\; \tilde{\cat{F}}_{\alpha\beta}=-\tilde{{\bf T}}_\alpha.\bm\gamma.\tilde{{\bf T}}_\beta$ and $\tilde{\cat{G}}^{\alpha\beta}$
its appropriate Green function. The corresponding $1$-form on the principle bundle is
$\tilde{{\bm\omega}}^\alpha=\tilde{\cat{G}}^{\alpha\beta}\tilde{{\bf T}}_\beta.\bm\gamma$ leading to the associated horizontal projector
$$\tilde{\Pi}^i_{\;\;j}(\chi)=\delta^i_{\;\;j}-\tilde{T}^i_{\;\;\alpha}\tilde{\omega}^\alpha_{\;\;j}.\;\;\blacksquare$$
An example of the form of the projectors for an extended light-cone gauge is given in \cite{MGW,LC08}.

\paragraph{\;{\it Recollection: from local to global convolution \newline}}

{\bf Theorem 3.3.1}{\;\it Global convolution}: Let $(\Omega_j)_{j\in {\bf J}}$ be a family of open subset of $\RR^{\;n}$  and  $(T_j)_{j\in {\bf J}}$ be a family of 
 distributions such that $T_j \in {\mathcal D}^\prime(\Omega_j)$ . The family $(T_j)_{j\in {\bf J}}$ is supposed to fulfil the compatibility 
 condition $T_{j\mid\Omega_j\cap \Omega_k}=T_{k\mid\Omega_j\cap \Omega_k}$ for all $j,k \in {\bf J}$. Then there exists a single distribution
  $T$ on $\Omega=\bigcup_{j\in {\bf J}}\Omega_j$ such that the restriction of $T$ to each $\Omega_j$ is $T_j$.\\
The proof is given in Appendix B of \cite{JPHYSA1}.\;\;$\blacksquare$\\
It follows that a unique recollection of the localised convoluted expression $\mathscr{A}^{\catj}\star_{\;_{\hskip -11.pt \mbox{loc}}}\!\!\!\!{\;}^{\perp}\!\rho(x)$ exists in $T_p^\star\Omega$, which is the global Lie-algebra-valued one form looked after.\\
\subsubsection{The Gaussian integrator on the space of the convoluted smooth vector fields}

From the convolution operation the space $\cat{g}\otimes T_p^\star\Omega$ of the global and smooth Lie-algebra-valued one form $\curlA=\mathscr{A}\star\rho$  inherits the topological properties of the Schwartz space  ${\mathcal S}$ where the test function $\rho$ lives. In particular a Gaussian integrator ${\mathcal D({\curlA})}$ exists, which under the projection $\tilde{\Pi}(\chi)\curlA=\curlA^{\perp}$ separates into

\begin{equation}\label{pathmeas}
{\mathcal D}({\curlA})={\mathcal D}(\bm \xi){\mathcal D}(\curlA^{\perp})(\mbox{det}{\bm\gamma}^{\perp})^{\frac{1}{2}}(\mbox{det}\tilde{\cat{G}})^{\frac{1}{2}}(\mbox{det}{\bf Q})\delta(\bm\chi),
\end{equation}
where both $(\mbox{det}{\bm\gamma}^{\perp})^{\frac{1}{2}}$ and $(\mbox{det}\tilde{\cat{G}})^{\frac{1}{2}}$ are independant of the gauge parameter $\bm \xi$ and the Gaussian integrator on the vertical component of ${\curlA}$ has been traded to that of the gauge parameter $\bm \xi$, for the space of orbits is also fixed by the gauge condition $\bm\chi=0$. This is just the Faddeev-Popov procedure \cite{FadPop}. As in the example of Sec.3.2 the finite Gaussian integration on the gauge parameter $\bm \xi$ factors  out and plays no physical r\^ole. Morover, even though ghosts are present through the gauge condition, with the Landau-DeWitt gauge prescription ghost vertices vanish for Yang-Mills theory\cite{DeWMol,ParTom}.\\
\section{ The functional integral for pure gauge theories}
\subsection{geometrical considerations about the  local gauge-fixing condition}
Since a {\it bona fide} gauge condition should define an hypersurface $\Sigma^\alpha$  which intercepts each orbit only once\footnote{It may not be possible to formulate such a gauge condition globally in non abelian gauge theories, but it does not arise from local considerations \cite{Singer}.},  points on $\Sigma^\alpha$ can only be connected by geometry conserving geodesic curves: $\Sigma^\alpha$ is dubbed as a {\it regular geodesic hypersurface}. Geometrically  the  intersection of any plane perpendicular to  a tangent plane to $\Sigma^\alpha$ at any of its point $p$  is a {\it smooth} curve on $\Sigma^\alpha$ with no critical point. $\Sigma^\alpha$ is a topological compact space with refined open coverings\footnote{For example the n-sphere $S^n$ is closed and bounded, that is compact, with a relative topology induced by the usual topology on $\RR^{n+1}$. }. Clearly $\Sigma^\alpha$ inherits its local topology from the convoluted nature of the gauge field components themselves. Any neighbourhood  on $\Sigma^\alpha$ of a point  $p$ is then described by a {\it regular} 3-dimensional one-form $\omega$, possibly defined  in terms of a local parametrisation of the surface. A chart for this part of the surface follows then by projection onto a tangent plane to  $\Sigma^\alpha$ at point $p$.\\
Apart from these considerations the gauge-fixing condition is only constrained by dimensional arguments \cite{Wein}. The fields $\curlA^{\alpha}_\mu$  all have dimensionalities $d=1$ (in power of mass)  so that $\bm\chi^\alpha$ is at most of dimensionality $d=2$, for ({\it cf} Eq.(\ref{convol})) $\int_{\zeta({\bf B})}D_{vol}({\bf B}) \bm\chi^\alpha\bm\chi_\alpha$ is dimensionless. \\
The Landau-DeWitt covariant gauge  \cite{FradTse} separates the gauge field into a sum of a background contribution ${\;}^{bg}\!\curlA^{\alpha}$ and of a quantum field ${\;}^{qt}\!\curlA ^{\alpha}$. The geometrical separation of these two contributions is accomplished with the projection operator introduced in subsection $3.3.2$. In the condensed notation the set of vector fields  that generate the fibre are here given by ${\bf T}^i_\beta\equiv\mathcal{K}^i_{\;\beta}$ and the projection operators write
\begin{eqnarray*}
\fl &\;\;&\bfsqcap^{i}_{\;j}=\delta^{i}_{\;j}-\bfsqcup^{i}_{\;j} \;\; \mbox{with} \;\; \bfsqcup^{i}_{\;j}=\mathcal{K}^{i}_{\;\tilde{\alpha}}\gamma^{\tilde{\alpha},\tilde{\beta}}\mathcal{K}_{\tilde{\beta},j}, \;\;  \gamma^{\tilde{\alpha},\tilde{\beta}}=[\gamma_{\tilde{\alpha},\tilde{\beta}}]^{-1} \;\; \mbox{and} \;\; \gamma_{\tilde{\alpha},\tilde{\beta}}=\mathcal{K}^{i}_{\;\tilde{\alpha}} g_{i,j} \mathcal{K}^{j}_{\;\tilde{\beta}}.\\
\fl &\;\;& \quad
\end{eqnarray*}
{\bf Proposition 4.1:} In the background field gauge with Landau-DeWitt condition the decomposition $\curlA^{i}={\;}^{bg}\!\curlA^{i}+{\;}^{qt}\!\curlA^{i}$ is fully consistent and such that ${\;}^{bg}\!\curlA^{i}={\sqcup \kern -.7em \sqcup}^{i}_{\;k}\curlA^{k}$ and ${\;}^{qt}\!\curlA^{i}={\sqcap \kern -.7em \sqcap}^{i}_{\;k}\curlA^{k}$.\\
Proof: The  total gauge field $\curlA^i$ is decomposed  as
\begin{eqnarray*}
\fl \;\;\;\;\curlA^i&=&{\;}^{bg}\!\curlA^{i}+{\;}^{qt}\!\curlA^{i}=\delta^{i}_{\;j}\curlA^{j}=[\bfsqcap^{i}_{\;j}+\bfsqcup^{i}_{\;j}]\curlA^{j},\quad \mbox{then}\\
\fl \bfsqcap^{i}_{\;j}\curlA^{j}&=&[\delta^{i}_{\;j}-{\sqcup \kern -.7em \sqcup}^{i}_{\;j}]{\;}^{qt}\!\curlA^{j}+\bfsqcap^{i}_{\;j}{\;}^{bg}\!\curlA^{j}
={\;}^{qt}\!\curlA^{i}-\mathcal{K}^{i}_{\;\tilde{\alpha}}\gamma^{\tilde{\alpha},\tilde{\beta}}\mathcal{K}_{\tilde{\beta},j}{\;}^{qt}\!\curlA^{j}+\bfsqcap^{i}_{\;j}{\;}^{bg}\!\curlA^{j}\\
\fl \;\;\;\;&=&{\;}^{qt}\!\curlA^{i}+ \bfsqcap^{i}_{\;j}{\;}^{bg}\!\curlA^{j} \;\; \mbox{for, from the gauge condition,}\;\; \mathcal{K}_{\tilde{\beta},j}{\;}^{qt}\!\curlA^{j}=0\\
\fl \;\;\;\; &=& {\;}^{qt}\!\curlA^{i} \;\; \mbox{if} \;\; \bfsqcap^{i}_{\;j}{\;}^{bg}\!\curlA^{j}=0 ,\;\; \mbox{which imposes}\;\; {\;}^{bg}\!\curlA^{j} \propto \bfsqcup^{j}_{\;k}\curlA^k, \;\;\mbox{for}\;\; \bfsqcap^{i}_{\;j}\bfsqcup^{j}_{\;k}=0. 
\end{eqnarray*}
 Moreover it is clear that the sum ${\;}^{bg}\!\curlA^{i}+{\;}^{qt}\!\curlA^{i}$ undergoes an ordinary gauge transformation, for the sum ${\sqcup \kern -.7em \sqcup}^{i}_{\;j}+{\sqcap \kern -.7em \sqcap}^{i}_{\;j}=\delta^{i}_{\;j}$ is gauge independant, although each individual projector is not.$\blacksquare$\\
The lesson from these arguments is clear: from the start it is always liable to use an horizontal test function  ${\;}^{\perp}\!\rho^{i}_{\;\;j}={\sqcap \kern -.7em \sqcap}^i_{\;\;j}\rho$ \; in Schwartz Space $\mathcal{S}(\bar{\RR}\!^4)$. Thereby the geometrical separation and the Landau-DeWitt gauge condition are directly enforced at the level of the classical Lagrangian. 
\subsection{The electromagnetic Abelian gauge group $U(1)$}
Here the projection operator takes the simple form
\begin{equation}\label{PHU1}
\;\; \bfsqcap_{\mu\;\nu}(x,y)=(g_{\mu\;\nu}-\frac{\partial_\mu\partial_\nu}{\partial^2})\delta^4(x-y).
\end{equation}
Picking up an horizontal test function accordingly it is straightforward to show \cite{MGW} that
only the relevant degrees of freedom of the convoluted quantum vector field enters the lagrangian density and moreover the gauge condition $\partial_\mu{\;}^{qt}\!\curlA^{\mu}=0$ becomes identically fulfilled.\\
The group space of $U(1)$ is the circle $S^1$ which, if parametrised as $(\xi^1)^2+(\xi^2)^2=1$, transforms to $r^2=1$ by the trivial ({\it Hopf}) map  defined by $S^1 \stackrel{\pi}{\mapsto} S^1$
$ \{\xi^1= r cos(\theta), \xi^2= r sin(\theta)\}.$ The complex plane is separated in its upper and lower halves defining the fibre bundle structure with transition function $\xi^1+\imath \xi^2 \in U(1)$. Hence the $U(1)$ fibre bundle is characterised
by the homotopy class 1 of $\pi_1(S^1)=\bfZ$, the additive group of integers. This Hopf-map may be understood as describing a magnetic monopole of unit strength \cite{RYD},\cite{MGW}(Appendix B2)\footnote{With the horizontal test function defined from
Eq.(\ref{PHU1}) the $U(1)$ phase factor involves ${\;}^{qt}\!\curlA^{\mu}$ only, hence bearing different magnetic charges in terms of the winding number (e.g Chern characters more generally).}
\subsection{the non-Abelian gauge groups $SU(2)$ and $SU(3)$}
 In this cases the Landau-DeWitt  gauge condition $\bm\chi^\alpha$ may be taken as
\begin{equation}\label{LDWgauge}
\bm\chi^\alpha=\partial\cdot{\;}^{qt}\!\curlA^{\alpha}+f^\alpha_{\;\;\beta\gamma}{\;}^{bg}\!\curlA^{\beta}\cdot{\;}^{qt}\!\curlA^{\gamma}\equiv{\;}_{\;\;x}^{\star}\nabla^{a}_{\;\;b\mu} {\;}^{qt}\!\curlA^{b\mu}(x),
\end{equation}
where, in the end part of this relation, the condensed notation for $\alpha$ and $\gamma$ is made explicit in terms of the respective Lie indices $a$ and $b$ and space variable $x$. For any field $\phi$ in the adjoint representation we have,
\begin{equation}\label{nabla}
{\;}^{\star}\nabla^\alpha_{\;\;\gamma\mu}\phi^\gamma\equiv{\;}_{\;\;x}\partial_\mu \phi^a(x)+f^{a}_{\;\;bc}{\;}^{bg}\!\curlA^{b}_{\;\;\mu}(x)\phi^c(x)
\end{equation}
It is seen that, under an infinitesimal gauge transformation $\xi,\;\;\bm\chi_\alpha$ transforms as${\;}^{qt}\!\curlA_{\alpha}$:
$$\delta\bm\chi_\alpha=-f_{\alpha\beta\gamma}\xi^\beta\bm\chi^\gamma.$$ 
Then the term in $\bm\chi^\alpha\bm\chi_\alpha$, introduced in the Lagrangian through
the usual exponential transcription of the $\delta$-functional form of the gauge condition, is invariant by antisymmetry of the structure constant $f$ in the interchange of indices
$$\delta\left[\bm\chi^\alpha \bm\chi_\alpha\right]=-2f_{\alpha\beta\gamma}\bm\chi^\alpha\xi^\beta\bm\chi^\gamma=0.$$\\
In the Landau-DeWitt gauge as expressed in Eq.(\ref{LDWgauge}) it is straightforward to find the expression of the matrix (sect. $7.3.1$ of \cite{ParTom}) $Q^\alpha_{\;\beta}$ from the functional derivative of $\bm\chi^\alpha$ with respect to the convoluted field variable ${\;}^{qt}\!\curlA^{j}\equiv{\;}^{qt}\!\curlA^{b\;\mu}(y)$
$$\bm\chi^\alpha_{,\;j}=\frac{\delta\bm\chi^a(x)}{\delta{\;}^{qt}\!\curlA^{b\mu}(y)}=
{\;}_{\;\;x}^{\;\star}\nabla^a_{\;\;b\mu}\;\delta^{(4)}(x-y),$$
where $x,y$ stands for the coordinate representations of points $p,q$ with $q$ in the neighbourhood of $p$. It follows that ({\it cf} Eqs.(\ref{convol},\ref{Kij}))
\begin{eqnarray}
Q^a_{\;b}(x,y)&=&\int_{\zeta({\bf B})}D_{vol}({\bf B}){\;}_{\;\;x}^{\;\star}\nabla^a_{\;\;c\mu}\;\delta^{(4)}(x-z){K^{c\mu}}_{b}(z,y),\nonumber\\
&=&-\int_{\zeta({\bf B})}D_{vol}({\bf B}){\;}_{\;\;x}^{\;\star}\nabla^a_{\;\;c\mu}\;\delta^{(4)}(x-z){{\;}_{\;\;z}^{\;}\nabla^{c\mu}}_{\;b}\;\rho(z-y),\nonumber\\
&=&-{\;}_{\;\;x}^{\;\star}\nabla^a_{\;\;c\mu}{{\;}_{\;\;x}^{\;}\nabla^{c\mu}}_{b}\;\rho(x-y)\label{Qab}
\end{eqnarray} 
In Yang-Mills theory the essential benefit of the Landau-DeWitt gauge is achieved if the background field is solution of the classical equations of motion, 
for all the terms in the effective action arising from the connection vanish \cite{ParTom,DeWittQFT}. In general the calculation of det$Q^\alpha_\beta$ is not an easy task, but becomes much simpler
for constant background fields which, for non-Abelian gauge theories, cannot be removed by a gauge transformation. In this case the ${\;}_{\;\;\,}^{\;\star}\nabla$'s can be diagonalised in a momentum space
 representation and with a constant background fields which, 
for non-Abelian gauge theories, cannot be removed by a gauge transformation. In this case the ${\;}_{\;\;\,}^{\;\star}\nabla$'s can be diagonalised in a momentum space representation and it may be shown  
that the complete  calculation of the effective action to one-loop (\cite{Wein} sect.$17.5$) can be performed  within the analysis of \cite{MGW}.\\
These constant fields settle the fibre $G_p$ at point $p={\sqcap \kern -.7em \sqcap}(u), u\in \mbox{P(M,G)}$ (principal bundle).
 With the element $\mathcal{K} \in \cat{g}$ a fundamental vector field $\Ksh \in T^{\parallel}P$ is generated by (\cite{Nakah}, chap.10)
$$\Ksh f(u)=\frac{d}{dt}f(u \exp(t\mathcal{K}))\mid_{t=0}$$
where $f\;:\; P \rightarrow \RR $ is an arbitrary smooth function. $u\exp(t\mathcal{K})$ defines a curve through $u$ in $P$ which lies within the fibre $G_p$. The vector $\Ksh$  is tangent to $G_p$ at $u$, hence $\Ksh$ is in the vertical subspace $T^\parallel P$ of the tangent space of $P$. The ${\#}$ operation preserves the Lie algebra structure $[(\Ksh)^\alpha,(\Ksh)^\beta]=[\mathcal{K}^{\alpha},\mathcal{K}^{\beta}]^{\#}$. In effect the ${\#}$ operation\footnote{Clearly the reference frame at point $p$ is unaffected by any change leading to  new non-zero constant background fields}  is then a displacement of the covariant derivative along the fibre at point $p$. The corresponding Lie group of $(\Ksh)^\alpha$ being parallelisable the tangent space at each point $p$ of the manifold is characterized as follows. Consider the norm
\begin{equation*}
{\;}_{\;\;\,}^{\;a}N^2=(\Ksh)^a_{\mu}{\;}(\Ksh)^{a,\mu},\quad\mbox{and}\quad x^a_\mu=\frac{(\Ksh)^{a}_{\mu}}{{\;}_{\;\;\,}^{\;a}N}.
\end{equation*} 
Then ${\bf X}^a=\{(x^a_1,x^a_2,x^a_3,x^a_4)\mid{\displaystyle \sum_{\mu=1}^4}(x^a_\mu)^2 =1\}.$
 The three orthogonal vectors 
$${\bf e}^a_1=(-x^a_2,x^a_1,-x^a_4,x^a_3),\quad {\bf e}^a_2=(-x^a_3,x^a_4,x^a_1,-x^a_2),\quad {\bf e}^a_3=(-x^a_4,-x^a_3,x^a_2,x^a_1)$$ 
are orthogonal to ${\bf X}^a=(x^a_1,x^a_2,x^a_3,x^a_4)$, linearly independent  and thus define the tangent space $T_X{\bf S}^3$. $X^a$ is then naturally the unit fundamental vector field tangent to the orbit at point $p$. This reference frame at point $p$ can  be parallel-transported further in a geometry-conserving way along the geodesic from the point $p$ to any point $q$ on the gauge surface.\\
We shall denote ${\;}^{\parallel}{\bf e}^{a}$ this unit fundamental vector field which can be identified to ${\bf X}^a$  and ${\;}^{\perp}{\bf e}^{a}$ the collection of unit vectors $\{{\bf e}^a_1,{\bf e}^a_2,{\bf e}^a_3\}$. Then the covariant  derivative in the direction of the orbit at point $p$ is
\begin{equation}\label{Devorb}
 \Xstarnabla^a_{\;c\mu}=({\;}^{\parallel}{\bf e}^{a}.{\;}^{\star}\nabla^a_{\;c}){\;}^{\parallel}{\bf e}^{a}_\mu .
\end{equation}
In effect picking up an horizontal test function puts the convoluted quantum field in the hyperplane specified by ${\;}^{\perp}{\bf e}^{a}=\{{\bf e}^a_1,{\bf e}^a_2,{\bf e}^a_3\}$, so that the action on ${\;}^{qt}\!\curlA^{a}$ of the covariant  derivative in the direction of the orbit at point $p$ defined by Eq.(\ref{Devorb})  is identically null.
Moreover the evaluation of $Q^a_b$ in Eq.(\ref{Qab}) reduces to
$$Q^a_{\;b}(x,y)= -[({\;}_{\;\;x}^{\;\star}\nabla)^2]^a_{\;b}\;\rho(x-y).$$ 
Then the calculation to one-loop with horizontal test functions follows \cite{Wein}.\\
Here the fibre is the unit quaternion $S3=SU(2)$. Hopf has shown that $S3$ is a $U(1)$ bundle over $S2$. The unit-sphere embedded in $\RR^4$ is such that $\Sigma_{i=1}^4 (x^i)^2=1$. The Hopf map $\pi : S^3 \to S^2$ is defined by
\begin{eqnarray*}
\zeta^1&=&2(x^1 x^3+x^2 x^4)\\
\zeta^2&=&2(x^2 x^x3- x^1 x^4)\\
\zeta^3&=&(x^1)^1+(x^2)^2-(x^3)^2-(x^4)^2,
\end{eqnarray*}
 and the transition function on the equator is $t_{NS}(\zeta)=\zeta^1+\imath \zeta^2 \in U(1)$:  the Hopf map is characterized by the homotopy class 1 of $\pi_3(S(U2))=\pi_1(U(1))=\bfZ$.\\
In going to  $SU(3)$, the symmetry of strong interactions, it is known \cite{Steen} that pairs of maps exist such that
$$SU(2)\stackrel{\rm {\bf \imath}}{\mapsto} SU(3) \stackrel{\rm {\bf \pi_3}}{\mapsto} S^5,$$
where $\imath$ is the canonical inclusion and $\pi_3$ maps a matrix $A$ to $A e_0$ with $e_0$ the unit vector with entries $0$ except $1$ for the last one. In effect  {\it up to isomorphism} $SU(3)$ {\it is the unique non-trivial} $SU(2)$-{\it bundle over the} $5${\it-sphere}\footnote{Locally constructed from spheres, but not globally}. Accordingly the full symmetry group of the standard model $G_{SM}=U(1)\times SU(2)\times SU(3)$ is locally \cite{AlSoco} $S^1\times (S^3)^2\times S^5$. Globally the characteristic function of $SU(3)$ is the suspension of the Hopf map $S3 \stackrel{\rm {\bf h}}{\mapsto} S^2$.
\section{Concluding remarks}
In analysing Cartier-DeWitt-Morette's treatment of functional integration we were driven to the observation that one of their step marks concerns the nature of fields  as  ${\mathcal C}^\infty$-objects, living in the Schwartz-space. In the present paper we have shown how this goal can be achieved with our Operator Valued Distribution approach of \cite{JPHYSA1,MGW}, via a convolution of the bare fields with ${\mathcal C}^\infty$-test functions. In effect the first outcome of this convolution is to bridge the gap between singular bare fields as distributions and the basic requirements of differential geometry where fields are regular objects living on differential manifolds. We focused next on the definition of the volume element in the space field variables. It is a fundamental issue for a mathematically well defined object in the sense of Measure Theory. The simple example of section (3.2) introduces  the important notion of the invariant Haussdorf measure in the space of fields, necessary for the definition  of the Laplace-Stieltjes transform. It also makes the link with the general geometrical treatment of pure gauge theories. To deal with the increasing interest in gauge theories in curved space-time we formulate the extension of the notion of convolution in such spaces. Most naturally it takes the form of parallel transportation of the test function by the Killing vector flow along the geodesics. Then we showed how the analysis in terms of connections on fibre bundles finds its counterpart in the functional integral formulation with test functions. The important point there is: the separation into physical and non-physical degrees of freedom can be done at the level of test functions (horizontal and vertical test functions with respect to the tangent space) through the action of horizontal or vertical projection operators. Via convolution this separation is implemented  directly into the Lagrangian. This is also true for the the Landau- DeWitt gauge condition which we use, for the implied separation into a background field and a quantum field is just enforced  through the action of the above projection operators. \\
In summary the Operator Valued Distribution approach via  convolution brings coherence into the interplay of differential geometry, topology (through a proper definition of a field space measure) and the formulation of Quantum Field Theory in terms functional integration. \\
It is an open question  whether or not the present conclusions would have implications in the joint approach of Parker and Toms \cite{ParTom}  to Yang-Mills and gravitation as gauge theories. \\
\par
\hspace*{0.5cm}{\bf Acknowledgements}\\
\par
We acknowledge the financial support from CNRS/IN2P3-Department during the course of
this work. EW is grateful to Denis Puy for his kind hospitality at the LUPM. We are indebted
to M. Slupinski (IRMA, Strasbourg) and C. Contou-Carr\'ere (L2C, Montpellier) for numerous
and valuable discussions.\\
\par
 
\clearpage
\appendix
\centerline{\bf Appendix }
\setcounter{equation}{0}
\section{Definitions and properties of the Laplace-Stieltjes transform}
For a quick out-look on Measures and Integration we refer  to \cite{CardeW} (1997 sect.$3$) and to \cite{Dud} for complete accounts on the subject. Here we give a little glossary on various definitions and properties found in the mathematical literature and used in the main text.
\subsection{Definitions and properties of measures. The Lebesgue-Stieltjes measure}
{\it $\sigma$-additive maps} Definition: Let $X$ be a set, ${\mathcal A}$ a family of subset of $X$ including $\sl0$ and $\mu: {\mathcal A} \to [0,+\infty]$ a  function of sets such that $\mu(\sl0)=0$. $\mu$ is dubbed $\sigma$-additive if for all $A \in {\mathcal A}$ and for all sequences of sets $A_n \in {\mathcal A} $ two-by-two disjoint  such that $A= \displaystyle{\cup_{_{\hskip -10.pt {n=1}}}^{^{\hskip -10.pt {n=\infty}}}} A_n$ it holds that $\mu(A)=\displaystyle{\Sigma_{_{\hskip -10.pt {n=1}}}^{^{^{\hskip -10.pt {n=\infty}}}}}\mu(A_n)$. \\
{\it $\sigma$-algebras} Definition: a $\sigma$-algebra  over a set $X$ is a family of subset of $X$ including $\sl0$ and $X$, stable by complementarity and by countable unions.\\
{\it Measurable sets} Definition: a measure is a $\sigma$-additive map on a $\sigma$-algebra of subset of a set $X$.
Then a measurable set is a couple $({\mathcal A},X)$ where ${\mathcal A}$ is a $\sigma$-algebra on $X$. A measured set is a triple $({\mathcal A},X,\mu)$ where $\mu$ is a measure on ${\mathcal A}$.\\
{\it Basic properties of measured sets} Let $({\mathcal A},X,\mu)$ be a measured set,\\
$(1)$ For all $A,B \in {\mathcal A}$, such that $A \subset B$, $\mu(A) \leq \mu(B)$,\\
$(2)$ For all $A \in {\mathcal A}$ and all sequences $A_n \in {\mathcal A}$ such that $A \subset \displaystyle{\cup_{_{\hskip -10.pt {n=1}}}^{^{\hskip -10.pt {n=\infty}}}} A_n$, $\mu(A) \leq \displaystyle{\Sigma_{_{\hskip -10.pt {n=1}}}^{^{^{\hskip -10.pt {n=\infty}}}}}\mu(A_n)$, \\
$(3)$ If $A_n \in {\mathcal A}$ is a growing sequence, of union ${A}$, $\lim \mu(A_n)=\mu(A)$,\\
$(4)$ If $A_n \in {\mathcal A}$ is a decreasing sequence, of intersection $A$, and if $\mu(A_1) < +\infty$, $\lim \mu(A_n)=\mu(A)$.\\
{\it Radon and Borel measures} Definitions:  if among measurable sets there are open and closed ones the corresponding measures are known as Radon measures. Let $({\mathcal A},X,\mu)$ be a measured set with $X$ a topological set. A Borel $\sigma-$algebra on $X$ is the $\sigma$-algebra  generated by open sets and named ${\mathcal B}(X)$ with elements dubbed {\it borelians of $X$}. A Borel measure on the topological set $X$ is a measure on  ${\mathcal B}(X)$. \\
{\it $\sigma$-lower-additive maps } Definition 1: let $X$ be a set, ${\mathcal A}$ a family of subset of $X$ and $\mu: {\mathcal A} \to [0,+\infty]$ a function such that $\mu(\sl0)=0$. One says that $\mu$ is $\sigma$-lower-additive if for all $A \in {\mathcal A}$ and for all sequences $A_n \in {\mathcal A} \; (n\in \bfN^*) \; A\subset \displaystyle{\cup_{_{\hskip -10.pt {n=1}}}^{^{\hskip -10.pt {n=\infty}}}} A_n \Rightarrow \mu(A) \leq  \displaystyle{\Sigma_{_{\hskip -10.pt {n=1}}}^{^{^{\hskip -10.pt {n=\infty}}}}}\mu(A_n)$. A $\sigma$-lower-additive map corresponds then to a growing sequence.\\
Defintion 2: an exterior measure is a $\sigma$-lower-additive function $\mu: {\mathcal P}(X)\to [0,+\infty]$.\\
{\it Canonical exterior-measure} Let $X$ be a set, ${\mathcal A}$ any family of subset of $X$ including $\sl0$ and a function $\mu: {\mathcal A} \to [0,+\infty]$ such that $\mu(\sl0)=0$. For $E\subset X$ one sets $\mu^*(E)=\mbox{inf}\{\displaystyle{\Sigma_{_{\hskip -10.pt {k=1}}}^{^{^{\hskip -10.pt {n=\infty}}}}}\mu(A_k)\;\mbox{with} \; A_k \in {\mathcal A}\; \mbox{and} \; E \subset \displaystyle{\cup_{_{\hskip -10.pt {k\geq 1}}}}A_k\}$, with the convention  $\mu^*(E)=\infty$ if such a sequence $A_k$ does not exist. Then:\\
(1) $\mu^*$ is an exterior measure,\\ 
(2) if $\mu$ is $\sigma$-lower-additive then $\mu^*$ extends $\mu$ and $\mu^*$ is dubbed {\it the canonical exterior measure} associated to $\mu$.\\
{\it Haussdorff measures and properties} Let $(X,d)$ be a metric set. The diameter of a non empty part $E$ of $X$ is
$\mbox{diam}(E)=\mbox{sup}\{d(x,y)\; \mbox{with}\;(x,y) \in E^2 \} \in [0,+\infty];\; d(\sl0)=0\;\mbox{ by convention}.$\\
Definition: Let $\varphi:\;\RR_+\to \RR_+$ be any function (dubbed the gauge) such that $\varphi(0)=0$  and $(X,d)$ a metric set. Let ${\mathcal A}_\epsilon$ be the family of subset of $X$ with diameter $\leq \epsilon$ and $\mu_\epsilon:\;{\mathcal A}_\epsilon \mapsto [0,+\infty]$ defined by $\mu_\epsilon(E)=\varphi(\mbox{diam}(E))$. Let ${\mathcal H}_\epsilon$ be the canonical exterior measure associated to $\mu_\epsilon$ such that ${\mathcal H}_\epsilon=\mbox{inf}\{\displaystyle{\Sigma_{_{\hskip -10.pt {n=1}}}^{^{^{\hskip -10.pt {n=\infty}}}}}\varphi(\mbox{diam}(E_n))\}$ where $(E_m)_{m\in \bfN^*}$ is a covering of $E$ with $\mbox{diam}(E_n)\leq\epsilon \; \forall n\in{\bfN^*}.$  
The function $\epsilon \to {\mathcal H}_\epsilon$ is decreasing. One sets ${\mathcal H}(E)\doteq\displaystyle{\lim_{\hskip -2.pt {\epsilon \to 0}}}{\mathcal H}_\epsilon(E)=\displaystyle{\mbox{sup}_{_{_{\hskip -14.pt {\epsilon > 0}}}}}{\mathcal H}_\epsilon(E).\; {\mathcal H}$ is an exterior measure on $X$ and a measure on ${\mathcal B}(X)$, dubbed {\it Haussdorff measure} for the distance $d$ and gauge $\varphi$.\\
Properties:
(1) Let $(X,d)$ and $(Y,\delta)$  be two metric sets and a function $\varphi: \RR^+ \to \RR^+$ such that $\varphi(0)=0$. Let ${\mathcal H}_X$ (resp. ${\mathcal H}_Y$)  be the Haussdorff measure on $X$ (resp. $Y$) for the distances $d$  and gauge $\varphi$ (resp. $\delta$ and gauge $\varphi$). If $f: X\to Y$ is an isometry, then for all $E\subset X$ one has $H_Y(f(E))=H_X(E)$. \\
(2) The Haussdorff measure on $\RR^n$ for the gauge $\varphi(r) = r^n$  and the norm ${\|.\|}_\infty$ is the translation invariant Lebesgue measure, for a translation preserving the metric is an isometry.\\
A Haussdorff measure can then be defined in Schwartz space  ${\mathcal S}$ of test functions $(f,g, etc...)$ of rapid decrease, for, besides the usual semi-norm, full norms with associated metric $d(f,g)$ can be defined, such as for example
 $$N_m(f)= \mbox{sup}_{x\in\RR^n}(1+\|x\|)^m\sum_{\mid\beta\mid\leq m}\mid\partial^\beta_x f(x)\mid \quad \forall m \in \bfN $$
 and the distance
$$d(f,g)=\sum_{m=0}^{\infty} 2^{-m}\mbox{min}\{1,N_m(f-g)\}$$.\\
{\it Lebesgue-Stieltjes measure}\\
 Definition 1: a part of $E \subset \Omega$ is called negligible (or   $\mu^*$-negligible)  if  $\mu^*(E)  =  0$. \\
Definition 2: a measure is called complete if all subsets of a negligible set are measurable.\\
 On a general ground the continuous functional $\Phi[\rho]$ Eq.(\ref{eq23}) defines a finite measure on ${\mathcal S}$ with density $\rho$. A unique and complete canonical measure is obtained from any measure by its extension to the $\sigma-$algebra generated by measurable and negligible sets.
The Lebesgue-Stieltjes measure, dubbed here $d\bfF$, is obtained from the Haussdorff measure  for the gauge $\bfF :{\mathcal S}\supset K_X \ni X \mapsto \bfF(X)=\int_{\RR^n}\One_{K_X}(Y)\rho(Y)dY $, where $\One_{K_X}$ is the characteristic function of the $\sigma-$algebra $K_X$\footnote{In one dimension the density is then just $\frac{d\bfF(x)}{dx}=\rho(x)$, for $\One_{K_x}(y)=\mbox{Heaviside}(x-y)$}.
\subsection{The Laplace-Stieltjes transform}
Let  $B\subset{\mathcal S}$  denotes the subset of real-valued functions $\bfF$ of strongly bounded variation over the $\sigma-$algebra $K_X$. The Laplace-Stieltjes transform is given by
$$\{{\mathcal L}^*\bfF\}(\bfs)= \int_{K_X}e^{-<\bfs,X>}d\bfF(X),$$
with $\bfs \in \CC$. This transform shares many properties of the usual Laplace transform (convolution, derivation ...). In particular
since $\bfF$ is a finite measure with density $\rho$ the Laplace-Stieltjes transform of $\bfF$ is the the Laplace transform of $\rho$
$$\{{\mathcal L}^*\bfF\}(\bfs)=\{{\mathcal L}\rho\}(\bfs)$$.
\section{Summary of the Cartier-DeWitt-Morette formulation of the path integral }
Given a $d$-dimensional paracompact differentiable manifold ${\mathcal M}^d$ the initial route in \cite{BCdeW,Lachap2} is defined on the space $\bfX$ of paths $x$ 
$$ \bfX \ni x:  \bfT \to {\mathcal M}^d, \quad  \bfT=[t_a,t_b],$$ and not on the space $\RR^d$ of discretized paths, for, as mentionned earlier, the definition of a measure when $d$ goes to infinity is problematic. The dual space $ \bfX^\prime$
is the space of linear forms $<x^\prime,x>_{\bfX}\quad\in \CC$. Assuming separability of
$\bfX^\prime$, complex Borel-measures $\mu,\Theta$ and $Z$ exist $\Theta$:
$\bfX\times \bfX^\prime \to \CC$ and $Z$: $\bfX^\prime \to \CC$. The Gaussian integrator ${\mathcal D}^s_{\Theta,Z}(x)$ on $\bfX$ is then such that
\begin{equation}\label{TZeq}
\int_{\bfX} \Theta(x,x^\prime;s){\mathcal D}^s_{\Theta,Z}(x)=Z(x^\prime;s),
\end{equation}
where $s$ is parameter: $s\in\{1,\imath\}$ and
$$\Theta(x,x^\prime;s)=\mbox{exp}\{-\frac{\pi}{s}\bfQ(x)-2\pi\imath <x^\prime,x>_{\bfX}\},$$
$$Z(x^\prime;s)=\mbox{exp}\{-\pi s W(x^\prime)\}.$$
Here $\bfQ$ is a non degenerate bilinear form on $\bfX$ such that $Re(\frac{{\bf{\mathcal Q}}}{s})>0$ and $W$ is its inverse on the dual space $\bfX^\prime$. A one parameter family of Gaussian integrator ${\mathcal D}\omega_s(x)$ is then introduced by
\begin{equation}\label{TZWeq}
\fl\int_{\bfX}e^{(-2\pi\imath<x^\prime,x>_{\mathbb{X}})}{\mathcal D}\omega_s(x):=\int_{\bfX}\mbox{exp}[-\frac{\pi}{s}\bfQ(x)-2\pi\imath <x^\prime,x>_{\mathbb{X}}]{\mathcal D}^s_{{\mathcal Q},W}(x):=e^{-\pi s W(x^\prime)}
\end{equation}
 Let $\Im(\bfX)$ be the space of functionals $F_\mu(x;s)$ relative to the Borel measure $\mu$ with
\begin{eqnarray}\label{FStr}
F_\mu(x;s)&:=&\mbox{exp}[-\frac{\pi}{s}\bfQ(x)]\int_{\bfX^\prime}\mbox{exp}[-2\pi\imath <x^\prime,x>_{\mathbb{X}}]d\mu(x^\prime)\\
& =&\Im[\mu](x)\mbox{exp}[-\frac{\pi}{s}\bfQ(x)],
\end{eqnarray}
where $\Im[\mu]$ is the Fourier-Stieltjes transform of the measure $\mu$ on $\bfX^\prime$. For the injective map $\mu \mapsto F_\mu \quad \Im(\bfX)$ is endowed with a norm $\|F_\mu\|$.\\
An integral operator $\int_{\Im(\bfX)}$ on $\Im(\bfX)$ is then defined by 
$$\int_{\Im(\bfX)}\Im[\mu](x){\mathcal D}\omega_s(x)=\int_{\bfX^\prime}\Im[\omega_s](x^\prime)d\mu(x^\prime).$$
$\int_{\Im(\bfX)}$ is a bounded linear form on $\Im(\bfX)$ with
$$\mid\int_{\Im(\bfX)}\Im[\mu](x){\mathcal D}\omega_s(x)\mid \leq \|F_\mu\|.$$ 
The generalisation to fields is based on the elimination of any dependence on the parameter $s$ present  in the different objects just defined. This is made in three steps.\\
-Step 1): Extend $s$ to $s \in \CC_{+}$ and make an affine transformation on $\bfT$
so that $\bfT \to [0,1]=:\mathbb{I}$. It leads to a map: $\bfX \ni x \mapsto x(t_b-t_a)^{-\frac{1}{2}}=:\tilde{x}$, with $x$ being $L^{2,1}$. From the duality
$$<x^\prime,x>_{\mathbb{X}}=\int_{\mathbb{T}}x^\prime(t)x(t)dt$$
it follows that $x^\prime \mapsto  (t_b-t_a)^{\frac{1}{2}}x^\prime$;\\
-Step 2):  The Gaussian functionals $\Theta$ and $Z$ are just such that the factor $(t_b-t_a)$ is eliminated by a redefinition of $s \to \tilde{s}$. In effect $\Theta(x,x^\prime;t_b-t_a)=\Theta(\tilde{x},\tilde{x}^\prime;1)$ and $Z(x^\prime;t_b-t_a)=Z(\tilde{x}^\prime;1)$;\\
-Step 3): To effectively remove the residual dependence on the parameter $s$ from the data $(\bfX,\Theta,Z)$ envisage $s$ as resulting from a localisation in an enlarged completed space $\bfX\otimes{\mathcal T}$ where ${\mathcal T}$ is a Banach space of pointed maps $t: (\mathbb{I},{\bf i}_{\bf 0}) \mapsto (\CC_{+},{\bf 0})$.\\
Then the definitions of Eqs.(\ref{TZeq},\ref{TZWeq}) can be extended to infinite dimensional $\bfX$  Banach spaces, e.g Schwartz spaces of fields $\varphi$  introduced in Eq.(\ref{eq23}). \\ 
The measure $\mu$ being of Borel type with  $\Theta(\varphi,\omega)$ and $Z(\omega)\quad \mu-$integrable continuous, bounded functionals, under a change of variables the transformations of $\int_{{\mathcal S}}{\mathcal D}_{\Theta,Z}(\varphi)$ and $\int_{{\mathcal S}^\prime}d\mu(\omega)$ are closely related.\\
{\bf Proposition 2.1.1} (change of variables,Sect.{\bf 2.2} of \cite{CardeW}(1997) and  Appendix {\bf B5} of \cite{JlaCh}) Let $\bfY$ be a separable Banach space and $M: \bfX \to \bfY$ and $R: \bfY^\prime \to \bfX^\prime$  be diffeomorphisms with respective derivative mappings $M^\prime_{(x)}: T_x\bfX \to T_y\bfY$ and alike for $R^\prime_{(y^\prime)}$. Under the change of variable $x \mapsto M(x)=y$\\
$$\int_{\bfY} F_\mu(\tilde{y}){\mathcal D}_{\bar{\Theta},\bar{Z}}(\tilde{y})=\int_{\bfX} F_\mu(M(x)){\mathcal D}_{\Theta,Z}(x),$$
where $\bar{\Theta}\circ(M \times R^{-1}):=\Theta$ and $\bar{Z} \circ R^{-1}:= Z$.\\
{\bf Corollary 2.1.1} If $M(\bfX)=\bfX$ and $M^\prime$ is nuclear and $F_\mu(x) \in \Im_R(\bfX)$, then
\begin{equation}\label{Det}
\int_{\bfY} F_\mu(y){\mathcal D}_{\bar{\Theta},\bar{Z}}(y)=\int_{\bfX} F_\mu(M(x))
\mbox{Det}[M^\prime_{(x)}]{\mathcal D}_{\bar{\Theta},\bar{Z}}(x).
\end{equation}
\setcounter{equation}{0}
\section{Nuclearity of the translation operator on ${\mathcal S}$ }
Let $(\Tau_h)_\star(\rho)$ be the push forward of elements  $\rho$  of the  Banach space ${\mathcal S}$ of fast decreasing functions of L. Schwartz by the translation map $\Tau_h:\RR^4\to\RR^4,(\Tau_h)(x)=x+h.\quad{\mathcal S}$ is endowed with an invertible positive quadratic form ${\mathcal Q}$ with its associated norm and dual counterpart in ${\mathcal S}^\prime$.\\
Introduce a basis $(e_n)_{n\geq1}$ of ${\mathcal S}$ orthonormal for the quadratic form ${\mathcal Q}$, that is ${\mathcal Q}(\sum_nt_ne_n)=\sum_nt_n^2$. An operator
${\bf T}$ in ${\mathcal S}$ has a matrix elements $(t_{mn})$ such that 
$$ {\bf T}e_n={\displaystyle \sum_{m}} e_{m}.t_{mn}.$$
{\bf Definition:} ${\bf T}$ is nuclear in the sense of Grothendieck \cite{Groten} if the series $\sum_n t_{nn}$ converges absolutely for every orthonormal basis
of ${\mathcal S}$.\\
Pick up first  an orthonormal basis on $\mbox{{\bf L}}^2(\RR^4)$ ({\it e.g.} build out of direct products of basic Hermite's functions \cite{Puig}
$\zeta_n(x)=\pi^{-1/4}((n-1)!)^{-1/2} e^{-x^2/2} H_{n-1}(
\sqrt{2}x),\in{\mathcal S}(\RR), n\in\bfN^*, x\in\RR$), then
$$(\tau_x)_{mn}=<e_m,\Tau_x e_n>=<{\mathcal F}(e_m),{\mathcal F}\Tau_x{\mathcal F}^{-1}{\mathcal F}(e_n)>=\int_{\RR^4}\frac{d^4\bfk}{(2\pi)^4}\mbox{e}^{\imath \bfk.\bfx}\tilde{{\bf e}}_{\bf m}(\bfk){\tilde{e}^*_n(\bfk)},$$
and $ \forall n \in \bfN^*$, $\forall x \in \RR^4\backslash\{{\bf 0}\}$
$$\mid(\tau_x)_{nn}\mid < \int_{\RR^4}\frac{d^4\bfk}{(2\pi)^4}\mid\tilde{{\bf e}}_{\bf n}(\bfk){\tilde{e}^*_n(\bfk)}\mid=\int_{\RR^4}d^4\bfy\mid e_n(\bfy)\mid^2=1.$$
Hence the series $\sum_n (\tau_x)_{nn}$ converges absolutely $\forall x \in \RR^4\backslash\{{\bf 0}\}$.
\section{The geometrical setting of gauge theories}
\subsection{Trivial principal fiber bundle} 
The notion of a {\it fibre bundle} is appropriate for local problems of differential geometry and gauge field theories in particular. For an extensive review {\it cf e.g.} \cite{Mayer,DanVial,Nakah}. A fibre bundle is, so to speak, a topological space looking locally like a direct product of two topological spaces.
Let ${\mathcal M}$ denote the four-dimensional space-time manifold  and ${\mathcal G}$  the group of all gauge transformations. Any change in gauge is represented by a smooth assignment of an element of the gauge group to any point of the space-time, that is a map: ${\mathcal M}\mapsto {\mathcal G}$. The graphs of these maps live in a product space ${\bf P}={\mathcal M}\times{\mathcal G}$. ${\mathcal G}$ acts on ${\bf P}$ : If $p\;\in\;{\bf P},\;p=(\bfx,u),\;\bfx\;\in {\mathcal M},\;u\in{\mathcal G}$.
For $a\;\in\;{\mathcal G}$ one defines $R_a :\;{\bf P}\;\mapsto\;{\bf P}$ by $R_a(p)=pa=(\bfx,ua)$. Suppose there is any $p\;\in\;{\mathcal G}$ such that $R_a(p)=p$, then $a=e$, the identity element of ${\mathcal G}$. Hence there is an equivalence relation between points of ${\bf P}$: $p \sim p^\prime$ {\it i.e.}:  $\exists\;a \in {\mathcal G}$ such that $p^\prime=p$. It is apparent that the equivalent classes
can be labeled by the points in ${\mathcal M}$, that is the quotient of ${\bf P}$ by the equivalence relation is just\footnote{{\it cf} footnote 5} ${\mathcal G}$. The equivalence relation induces a canonical projection 
$$\Pi\;:\{\bfx\;\in {\mathcal M},\;u\in{\mathcal G},\;{\bf P}\; \stackrel{\rm \Pi}{\rightarrow}\; {\bf P}|\;\Pi(\bfx,u)=\bfx \},$$
whereby two equivalent points,$(\bfx,u)$ and $(\bfx,ua)$, project to the same point $\bfx \in {\mathcal M}$.$\;\Pi^{-1}(\bfx)$ is called a {\it fibre}, isomorphic to ${\mathcal G}$. The triple $\{ {\bf P},{\mathcal G},\Pi\}$ forms a {\it trivial principal fibre bundle} over the base space ${\mathcal M}$ with structure group ${\mathcal G}$ and projection $\Pi$.
\subsection{Connections on principal fibre bundles} We adopt here the presentation of
\cite{Nakah}. In line with our simple example of section (3.2)
 the approach is based on the separation of the tangent space $T_p{\bf P}$ into `vertical' and `horizontal' sub-spaces. Let ${\mathcal G}$ be a Lie group. The right action $R_u$ on elements of ${\mathcal G}$ is as defined earlier and the left action is such that $L_u(a)=au$ for $\;u,a \in {\mathcal G}$. $L_u$ induces a map $L_{u^\star}: T_a({\mathcal G})\;\mapsto\;T_{au}({\mathcal G})$. Left invariant vector fields $\bf A$ are such that $L_{u^\star}{\bf A}|_a={\bf A}|_{au}$. They form a Lie algebra of ${\mathcal G}$, dubbed $\cat{g}$, with a vector space isomorphism $\cat{g}\simeq T_e({\mathcal G})$ and $e$ the unit element. $\cat{g}$ closes under the Lie bracket $\left[{\bf T}_\alpha,{\bf T_\beta}\right]=f_{\alpha\beta}^\gamma{\bf T}_\gamma$, where $\{{\bf T}_\alpha\}$ is the set of generators of $\cat{g}$ and $f_{\alpha\beta}^\gamma$ the group structure constants. A connection on ${\bf P}$ is a unique separation of the tangent space $T_p{\bf P}$ into the vertical 
and horizontal sub-spaces $T^{\parallel}_p{\bf P}$ and $T^{\perp}_p{\bf P}$ according to the definition $3.3.1$  and with  projection operators built from the $1$-form on the principal bundle. Thereby the link is established with the covariant derivative specified in terms of the Riemannian  connection $\bm\Gamma_\gamma$ (or extension of it, whatsoever).

\end{document}